\documentclass[journal]{IEEEtran}

\usepackage{amsthm}
\usepackage{mathrsfs}
\usepackage{cite}
\usepackage{graphicx,amssymb,lineno}
\usepackage{amsmath,amsfonts,amssymb}

\newtheorem{definition}{Definition}
\usepackage{algorithm}
\usepackage{algorithmic}
\usepackage[usenames]{color}
\usepackage{float}
\usepackage{multirow}
\usepackage{mathtools}
\usepackage{cite}
\usepackage{booktabs}

\usepackage{amsmath}

\renewcommand{\algorithmicrequire}{\textbf{Initialization:}}

\usepackage{graphicx,graphics,color,epsfig,subfigure,graphpap,rotate}
\usepackage{times, verbatim, subfigure, epsfig, graphicx, latexsym, amsmath}
\usepackage{url}
\usepackage{subfigure}

\begin{document}

\title{Resource Allocation and Computation Offloading in a Millimeter-Wave Train-Ground Network}

\author{Linqian~Li,
        Yong~Niu,~\IEEEmembership{Member,~IEEE},
        Shiwen~Mao,~\IEEEmembership{Fellow,~IEEE},
        Bo~Ai,~\IEEEmembership{Senior Member,~IEEE},
        Zhangdui~Zhong,~\IEEEmembership{Senior Member,~IEEE},
        Ning~Wang,~\IEEEmembership{Member,~IEEE},
        and Yali~Chen

}

\maketitle

\begin{abstract}
In this paper, we consider
an mmWave-based
train-ground communication
system in the high-speed railway (HSR)
scenario, where the computation tasks
of users can be partially offloaded to the rail-side base station (BS) or
the mobile relays (MRs) deployed on
the roof of
the train.
The MRs operate in the full-duplex (FD) mode to achieve high spectrum utilization.
We formulate the problem of minimizing the average task execution latency of all users, under
local
device
and MRs energy consumption constraints.
We propose
a
joint resource allocation and computation offloading scheme (JRACO)
to solve the problem. It consists of
a
resource allocation and computation offloading (RACO) algorithm and
an
MR Energy constraint algorithm. RACO utilizes
the
matching game theory
to iterate between two subproblems, i.e.,
data segmentation
and user association and sub-channel allocation.
With the RACO
results,
the
MR energy constraint algorithm
ensures that the MR energy consumption
constraint is satisfied.
Extensive simulations
validate that
JRACO can effectively reduce the average latency
and
increase the number of served users compared with three baseline
schemes.
\end{abstract}


\section{Introduction}\label{S1}

With the explosive growth of
mobile computing applications,
the
5G and beyond
network systems
should be able to
provide
differentiated services
to various applications, with respect to
throughput, delay, and other performance indicators.
High-speed railway (HSR), as a convenient and green public transportation system, has been
developed
rapidly, and will become the future trend of global railway transportation
in many countries.
On the other hand, considering that users tend to cluster in a
railway cabin
and travel for long journeys, it is not inconceivable that they would be very interested in having Internet service
in the train, especially in multimedia services.
It is
thus
important to provide high quality broadband wireless access
for passengers.
However, the current communication technology, i.e., Global System for Mobile Communications-Railway (GSM-R), which has been widely used in high-speed railway scenarios, has
many
shortcomings, such as insufficient capacity, low network
resource
utilization,
and limited data service support \cite{LTE}.
When
the speeds of the train are over 300 km/h, the wireless
channel exhibits rapidly time-varying and nonstationary
features.
Accordingly, even the latest generation of HSR communications systems, LTE for Rail (LTE-R), cannot provide every user with broadband services due to
bandwidth limitations. The increasing demand for HSR communications leads to significant attention on the study
of spectrum extension.

To this end, millimeter wave (mmWave)
communications can provide transmission rates on the order of
gigabits for broadband multimedia services, including
high-speed
data transmission between devices, high-definition TV live broadcast, and cellular access, etc.~\cite{Gbit,3GPP}.
However,
mmWave signals
experience considerably
higher propagation losses
than sub-6GHz signals,
are
unable to penetrate most solid materials, and
are
particularly sensitive to blockage, resulting in higher signal attenuation and reflection~\cite{mmW1,mmW}.

Communication scenarios in
the HSR
environment mainly include intra-compartment communications
and train-ground communications.
In order to offer
broadband services, different reliable communication systems
that can provide better performance and mobility
support
for users can be
deployed
inside train
cabins,
such as wireless local area networks (WLAN), most
modern communication
devices
can use this system
if equipped with a WLAN network interface card~\cite{train}.
In addition,
mmWave communications can be leveraged
for train-ground communications, where mobile relays (MRs) can be deployed on the rooftop of the train.
In order to compensate for the severe attenuation of
mmWave signals,
directional antennas are usually
used to
achieve a
high antenna gain.

Another relevant technology is Full-duplex (FD) transmission,
which
has attracted great attentions in both academia and industry.
The FD
technology allows
wireless communication devices to
transmit
and receive signals simultaneously on the same frequency band by utilizing separated or shared antenna configuration. 
With the advances of self-interference (SI) cancellation techniques, FD communications can great enhance the spectrum utilization and system capacity, and
has been
recognized
as one of the key physical layer technologies of 5G
and beyond.

The biggest challenge of FD communications
is the elimination of
SI~\cite{fd1}. In order to achieve
high spectral efficiency,
existing
schemes
have been able to achieve
100dB reduction of SI through antenna separation, digital domain elimination, and analog domain elimination~\cite{fd100}, which
enables practical applications of
the
FD technology.
Considering the
HSR
scenario,
the MRs
could
adopt
FD transmissions
to simultaneously serve users and connect
to
the ground BS.
How to effective associate users to the MRs or the BS should be carefully determined.
When there are
many users, it is also a key problem of how to select
the set of users to serve to optimize
users' experience.

Recently, mobile edge computing (MEC) has emerged as a promising paradigm to
support many 5G and beyond applications
including latency sensitive services~\cite{mec, mec1,mec2}.
MEC can reduce the load on the core network and the data transmission delay by deploying nodes with computation processing capacity at the edge of
network to be closer to
users.
However, there has been very limited prior work on combing MEC and train-ground communication system in
the
HSR scenario.
Utilizing
mmWave
communications for train-ground communications
and
deploying MEC servers on high-speed trains, high-speed data transmission
will be enabled
with the help of FD MRs, which can significantly improve the broadband wireless communication service performance of the entire train-ground communication system.
The computing tasks of users can be
executed
locally, or be
offloaded to the MRs and
executed on the train, or be offloaded to the rail-side BS to be executed there.

In this paper, we
investigate
the problem of joint
optimization of partial computation offloading, user association, and resource allocation in
an mmWave FD
train-ground communications system.
Our objective is to
minimize the average delay for all users under the
energy consumption
constraints
of users and MRs.
In particular, each user can flexibly choose to partially offload its tasks
to
an MR
or the BS (in this case, the data transmission is still via the FD MRs).
Then
the
MRs and
the
BS can cooperate with each other to
execute the offloaded
tasks of users by
sharing
their limited computing resources.
The challenge is
how
to strike
a
balance between local execution and
offloaded execution
latencies
considering
dynamic data segmentation, distributed computing capacities, and FD transmissions.
We formulate a mixed integer nonlinear programming (MINLP) problem and propose a low complexity heuristic
algorithm, which solves the formulated problem by decomposing it into data segmentation, resource allocation problems for known user association and MR energy consumption control problem.

The main contributions made in this paper are summarized as follows.
\begin{itemize}
\item We propose
an MEC framework for
the mmWave
train-ground communication system, in which MRs
are deployed on the train to relay data between users and the rail-side BS and operating in the FD mode.
Both the BS and MRs serve as MEC servers.
Then, we
formulate the problem of
joint
user association, partial offloading, and resource allocation, aiming
to minimize the average latency
by taking account of the MRs energy consumption.

\item We propose a resource allocation and computation offloading (RACO) algorithm by decomposing the original problem into subproblems:
the subproblem
of data segmentation is solved by functional analysis, and
the user association and resource allocation subproblem
is solved by a matching game.
In addition, a heuristic algorithm is proposed to
enforce the
MRs energy consumption
constraint. 
Resource surplus and resource deficit scenarios
are all
considered.

\item We perform extensive
simulations
under various system parameter
settings
to validate the performance of our proposed scheme and compare
it with
four
benchmark schemes.
Our results validate that
the average latency
of all users
can be significantly reduced by the proposed scheme,
while keeping the MRs' energy consumption within constraints.
\end{itemize}

The
remainder
of the paper is organized as follows. In Section~\ref{S2}, we provide
an
overview of
related work. In Section~\ref{S3}, we introduce the train-ground communication system model,
including the
device, task, and partial offloading models. We discuss
the formulation of average latency minimization problem in Section~\ref{S4} and problem decomposition in
Section~\ref{S5}.
We
present
our proposed RACO and
MRs energy constraint algorithms in Section~\ref{S6}.
Our simulation study is presented in Section~\ref{S7}.
Finally, Section~\ref{S8} concludes this paper.

\section{Related Work}\label{S2}

There has been considerable interest in
FD communications
in cellular networks, mmWave networks, and heterogeneous networks
Wen \emph{et al.}~\cite{FD1} investigated the resource allocation and user scheduling problems to maximize
network throughput in a time-division cellular network, where FD was
adopted at
the BS and user equipments
still operate
in the HD mode.
To guarantee the
QoS requirements of
traffic
flows,
Ding \emph{et al.}~\cite{FD2} proposed
an
FD scheduling algorithm in the mmWave wireless backhaul
to
maximize
the number of completed flows.
Liu \emph{et al.}~\cite{liu} proposed a novel MEC framework with a user virtualization scheme in
a
software-defined network virtualization cellular network, and
introduced
user virtualization assisted by
FD
communications.
Lan \emph{et al.}~\cite{Lan} integrated FD
in
an
MEC enabled
HetNet.
A
maximization optimization problem of users' revenues is formulated, in which uplink FD transmissions
are
considered.

For the simpler
case of single-user systems,
in~\cite{Ning}, the cooperation of cloud computing and MEC
was
investigated
in the IoT setting,
and the single user computation offloading problem
was
solved by the
branch-and-bound
algorithm.  Kuang \emph{et al.}~\cite{Kuang} investigated the joint problem of partial offloading scheduling and resource allocation with multiple independent tasks. The goal
was
to minimize the weighted sum of the execution delay and energy consumption,
while guaranteeing the
transmit
power constraint of the tasks. The problem is solved by a two-level alternation method framework based on Lagrangian dual decomposition.

Other works
considered the
multiuser partial offloading MEC scenario. In~\cite{multi-a}, the
problem of
collaborative MEC offloading
for
multiuser multi-MEC in 5G HetNets
was
studied, and a game-theoretical computation offloading scheme
was
proposed.
Mao \emph{et al.}~\cite{multi-b} investigated
joint radio and computational resource management in
the
multi-user single-MEC scenario, with the objective to minimize the long-term average weighted sum power consumption of the MDs and the MEC server. An online algorithm
was proposed
based on Lyapunov optimization
for reduced power consumption.
In~\cite{multi-c}, Saleem \emph{et al.} jointly
considered
partial offloading and resource allocation to minimize the sum latency with energy efficiency for multi-user MEC offloading.
An expression to determine
the
optimal offloading fraction
was
derived such that energy consumed for local execution
would not
exceed the desired limit.
Chen \emph{et al.}~\cite{yali} minimized the total energy consumption of all users within the required latency.
User association
was jointly considered with
sub-channel allocation,
which were transformed into
a
two-sided matching game representing the resource competition among users.
Saleem \emph{et al.}~\cite{D2D}
focused
on minimizing the sum of task execution latency of all the devices in a shared spectrum
under
interference. Where desired energy consumption, partial offloading, and resource allocation constraints
were
considered.
A decomposition approach
was adopted
to solve the problem,
which iteratively reduced the parallel processing delay by adjusting
data segmentation and solving the underlying key challenge of interference in a shared spectrum.  However, the focus of the work
was
on interference management, and the iterative convergence speed of adjusting the unloading rate
was
relatively low.

As multiple users compete for
the
finite radio and edge computing resources,
some prior
works
deal
with resource allocation from
a
game-theoretic
perspective.
In~\cite{game1}, the authors
modeled
the network as a competitive game,
where users shared the communication channel to offload their computations.
Optimal
offloading decision
was derived
for minimizing user energy consumption while satisfying the hard deadline of the applications.
In~\cite{game2}, the problem of cloud-MEC collaborative computation offloading
was
investigated, and a game-theoretic collaborative computation offloading scheme was
proposed.
Di \emph{et al.}~\cite{game3} maximized the weighted total sum-rate by power control and sub-channel allocation, which was equivalent to a many-to-many matching game
where
peer effects existed.

Although great advances have been made,
the
problem of multiuser partial offloading in
the mmWave
band with FD communications
has not been fully
addressed in prior works so far.
Meanwhile, some previous studies sought to optimize
either
the computation offloading strategy or
computing resource allocation, but without considering both goals.
Motivated by
the related works,
we propose to
jointly optimize
computation offloading
and resource allocation,
by
modeling
the
sub-channel allocation problem as a matching game.
In addition, different from
the
previous works,
we consider two offloading locations, namely on the MR side and the BS side,
while
they communicate with each other in
the
FD mode
and cooperate with each other to complete users' computing tasks.
This paper also
considers the constraints on
the user
energy consumption
and the edge server on
the
MR
when minimizing the
system latency, which
has
not been fully
studied in prior works.

\section{System Overview}\label{S3}

\subsection{System Model}\label{S3-1}

\begin{figure}[!t]
\begin{center}
\includegraphics*[width=1\columnwidth,height=2.8in]{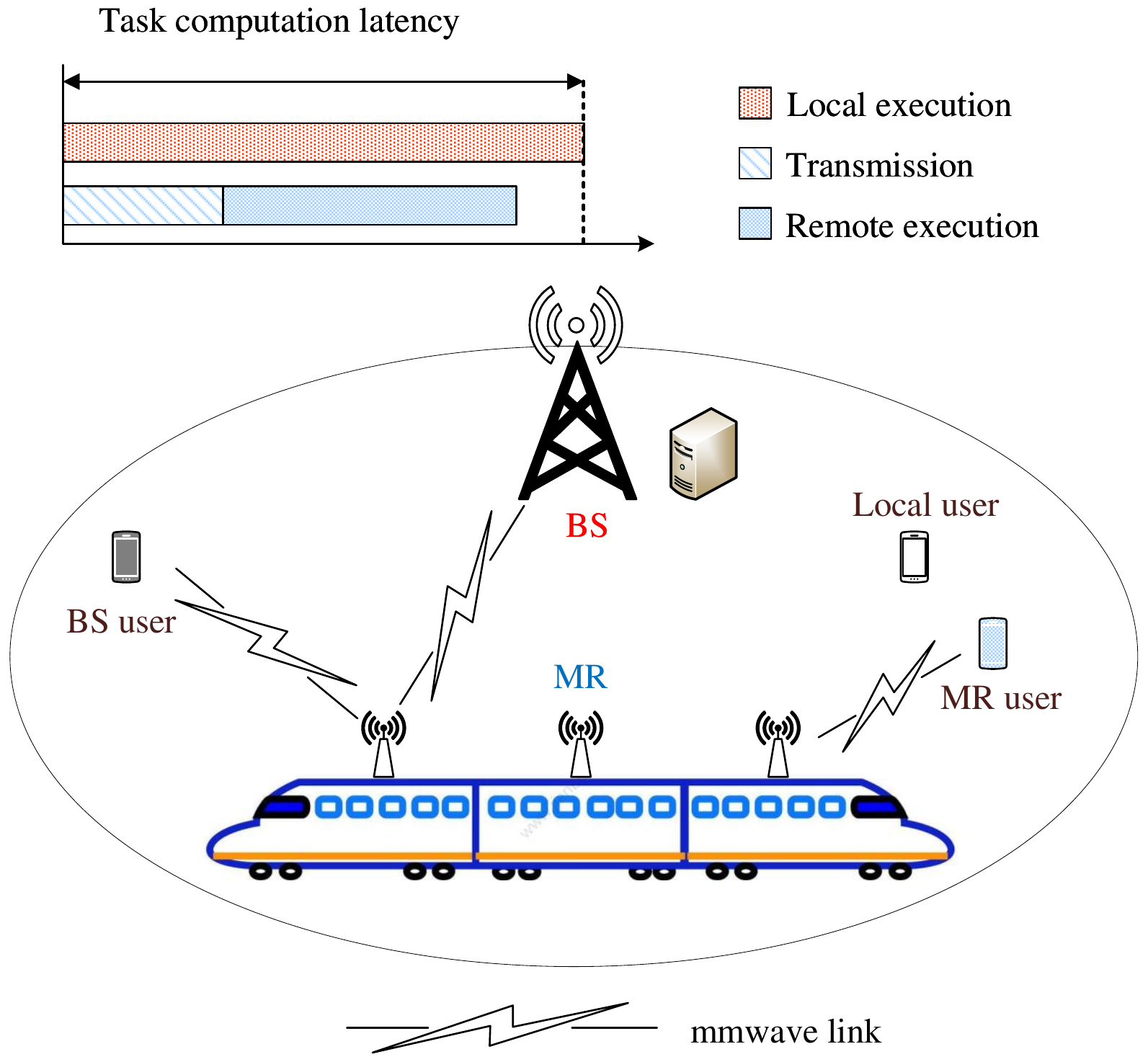}
\end{center}
\caption{Illustration of the mmWave train-ground communication system.}
\label{fig3.1}
\end{figure}

We consider
an mmWave-based
train-ground communication system using FD MRs
to serve multiple users, as shown in Fig.~\ref{fig3.1}.
There
are
one track-side BS and multiple MRs deployed on
the train.
The BS is installed with
an
MEC server and connects to the core
network.
The MRs
operate
in the FD mode
and
connect to users and the BS via wireless
mmWave
links.
Thus, the computing tasks of
users
can be offloaded to
the
MRs
to be
completed by the server on the train, or
to be
offloaded to the BS by the MRs.
We do not consider
the case
where
users are directly connected to the
track-side
BS, because users are often closer to
the
MRs
in practical scenarios. In addition,
the propagation of mmWave signals
is highly
vulnerable to various blockages
(e.g., the cabin wall or glass window)
due to its weak diffraction capability.
Thus, it is inappropriate for users to directly associate with the
BS.
All the devices in this system work in the
mmWave
band.
The BS and
MRs are all
equipped with steerable directional antennas to achieve
a high
antenna gain.

We assume that the equal bandwidth resources
allocated to
the MRs
are mutually independent.
At each
MR, multiple sub-channels are equally partitioned based on
the
available channel bandwidth. Different sub-channels adopt orthogonal frequency division and each sub-channel can serve at most one user.
Based on the above assumptions, there is no co-channel interference between users.
We assume that each user is aware of the location of
the
neighboring MR, as well as
its own location.
The user will establish an association
with the closest
MR.
Each MR has $S$ independent sub-channels of
equal
bandwidth,
denoted by
${\mathop\textbf{S}\nolimits}  = \{ 1,2, ..., S\}$.

Assume that there are $M$ users under the coverage of
an
MR.
The corresponding set of users is denoted as
${\mathop\textbf{M}\nolimits}  = \{ 1,2, ..., M\}$.
Users that can ultimately perform task offloading are divided into two categories, namely, BS users and MR users. Specifically,
the user set associated with
an
MR is denoted as $\mathcal {U}^R$, and the user set connected to
BS is denoted as $\mathcal {U}^B$.
The devices in the set have limited computation resources
but need to perform a delay-sensitive
and
computation-intensive task.
Specifically, we focus on
applications with partitionable data,
in which the amount of data to be processed is
known beforehand and the execution can be
in parallel.
That is,
the application
data can be partitioned into subsets of any size. In practice, many mobile applications are composed of multiple procedures,
making it possible to implement partial offloading.
We consider
the case where
each user has only one task to offload,
and characterize the task of user $m\in\textbf{M}$ with two key parameters $({d_m},{c_m})$, where ${d_m}$ is the data size of the task in bits,
and
${c_m}$ is the computation resource required to
process
one data
bit in CPU cycles per bit.
The quasistatic scenario
is considered in this paper,
where the set of users remains unchanged during
an
offloading period.
Table~\ref{table1} summarizes the
main notation used
in this paper.

\begin{table}[t]
\begin{center}
\caption{Notation}
\begin{tabular}{ll}
\toprule 
Notation & Description \\
\midrule 
$\textbf{M}$ & the set of users \\
$\textbf{S}$ & the set of sub-channels \\
$\textbf{Y}_m$ & the remotely
executed portion
set of user $m$\\
$\mathcal {U}^R$ & the set of MR users\\
$\mathcal {U}^B$ & the set of BS users\\
$d_m$ & the task data size of user $m$\\
$c_m$ & the task processing density of user $m$\\
$\lambda _m$ & the local
execution fraction of user $m$\\
$E_m$ & the energy constraint of user $m$\\
$E^R$ & the energy constraint of the MRs\\
$x_{ms}$ & whether sub-channel $s$ is occupied by user $m$\\
$W$ & the subcarrier bandwidth\\
$\beta$ & the SI cancelation level of the MRs\\
$f_m^R,f_m^B$ & the computation resource allocation of an MR, BS\\
$P_m^R$ &
the transmit
power of the
MR assigned to user $m$\\
\bottomrule
\end{tabular}
\label{table1}
\end{center}
\end{table}

\subsection{Partial Offloading Model}

We
consider the
partial offloading model, motivated by the fact that it benefits from parallel computing by efficiently utilizing the local and remote resources simultaneously~\cite{off}. In particular, we adopt
a
data-partition model,
where the input bits of the task can be arbitrarily divided due to bit-wise independence~\cite{bitwise}.
Under this model,
a fraction of
the
task
can be
processed locally,
and
the rest
can be
offloaded to the MR or BS. We introduce the parameter ${\lambda _m} \in \left[ {0,1} \right]$
to represent the ratio of the
locally executed portion
of user
$m$'s
task.
After determining the
partitioning of task data,
${{\lambda _m}{d_m}}$ bits
will be
processed locally, while ${{(1-\lambda _m)}{d_m}}$ bits
will be
offloaded to either the
BS or MR.
For ease of notation,
the set of binary variables $\textbf{Y}_m$ representing the remote execution locations
is introduced and defined as ${\mathop{\textbf{Y}_m}} = \{(\textrm{y}_m^R,\textrm{y}_m^B)|m\in\textbf{M}\}$,
where
${\rm{y}}_m^R = 1$ indicates that user $m$ offloads task to the MR, and
${\rm{y}}_m^B = 1$ indicates that user $m$ offloads task to the BS.

\begin{figure}[!t]
\begin{center}
\includegraphics*[width=0.8\columnwidth,height=1in]{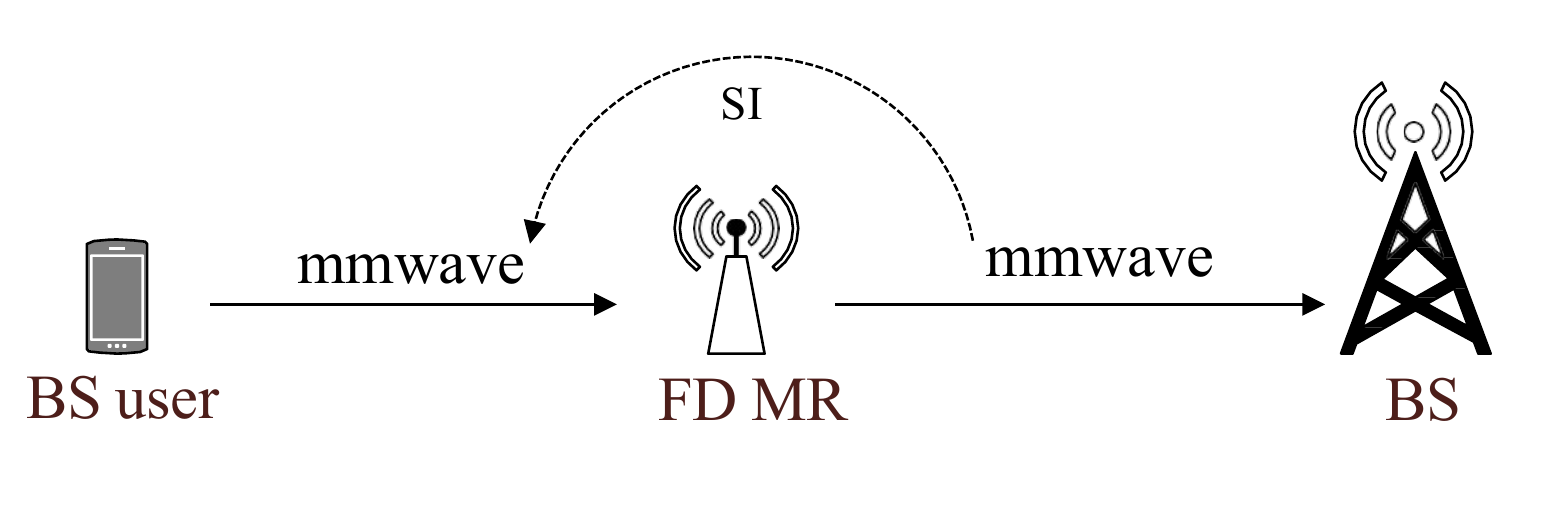}
\end{center}
\caption{Illustration of the self-interference (SI) at an FD MR.}
\label{fig3.2}
\end{figure}

For a user,
offloading tasks to
the
MR
or
BS are different in
the
wireless transmission
rate, because offloading tasks to
the
BS still
requires the help of MR, which works in
the
FD communication mode to
relay
the user's computing tasks to
the
BS. Since data is received and transmitted simultaneously in
the
FD mode, self-interference (SI) is introduced at the MR, as shown in Fig.~\ref{fig3.2}, which needs to be considered.

The execution
of a
user
task involves local execution, BS execution, and MR execution, which are modeled in the following.

\smallskip
\subsubsection{Local Execution}

We denote $f_m^L$ as the CPU
speed of completing
locally executed
tasks for
user $m$, which
is
measured by the
amount
of CPU cycles per second.
The advanced dynamic frequency and voltage scaling (DFVS) technique is adopted, which allows stepping-up or -down of the CPU cycle frequency. In practice, the value of  $f_m^L$  is bounded by a maximum value
$f_m^{max}$, which
is due to
the limitation of the user's computation capability.
The
user will process a fraction of its task locally, whereas the time consumption of local computation depends on the CPU clock frequency $f_m^L$, the data size of the task ${d_m}$,
and the number of CPU cycles required per bit $c_m$.
Then the local computation latency $t_m^L$ is given
by
\begin{align}\label{eqtl}
t_m^L = \frac{{{\lambda _m}{d_m}{c_m}}}{{f_m^L}}.
\end{align}

We model the local CPU power consumption
by
$\mu {(f_m^L)^3}$ as in \cite{mec1}\cite{off}, where $\mu$ is a coefficient depending on the chip architecture.
The local energy consumption for user
$m$
is shown as follows.
\begin{align}\label{eqEl}
E_m^L = \mu {(f_m^L)^3}t_m^L = \mu {\lambda _m}{d_m}{c_m}{(f_m^L)^2}.
\end{align}

\smallskip
\subsubsection{MR Execution}

 If the
 remote
 execution location of
 user $m$ is chosen
 to be the nearest MR, we need to determine firstly which sub-channel of this MR to access
 for offloading.
 For ease of notation, the binary variable ${x_{ms}}$ is defined,
where ${x_{ms}}=1$
indicates that
user $m$ occupies
sub-channel $s$.
Given the allocated spectrum resources,
the signal to interference plus noise ratio (SINR) of user $m$ connected
to MR
through
sub-channel $s$ can be expressed as
\begin{align}\label{eqsinr}
\Gamma_{ms}^R = \frac{{{H_s}{G_t}(m,R,s){G_r}(m,R,s)l_{mR}^{ - \alpha }{P_m}}}{{{N_0}W}},
\end{align}
where $H_s$ is the
gain of
sub-channel $s$
following the
Nakagami-$m_s$ distribution with
parameters $\{m_s, {w_s}\}$,
where
$m_s$ is the fading depth parameter and $w_s$ is the average power in the fading signal \cite{yali};
${G_t}(m,R,s)$ and ${G_r}(m,R,s)$ are the transmitting gain and receiving gain of
the
directional antenna pointing from user $m$ to
the
MR  on sub-channel $s$, respectively;
$l_{mR}^{ - \alpha }$ is the path loss of link $m$ to
the
MR  with path loss exponent
$\alpha$;
$P_m$ is the
transmit
power of user $m$ (to simplify
analysis, we
consider constant transmit power);
$N_0$ is the white Gaussian noise power spectral density;
and $W$ is the subcarrier bandwidth depending on channel bandwidth and the number of sub-channels partitioned. 
According to the Shannon theory, the achievable
transmission rate is
\begin{align}\label{Rsinr}
R_{ms}^R=W\log_2(1+\Gamma _{ms}^R).
\end{align}

Thus, the time consumed to upload $(1-\lambda_m)d_m$ data bits to
the
MR is
\begin{align}\label{eqtr}
t_m^R = \frac{{(1 - {\lambda _m}){d_m}}}{{R_{ms}^R}},
\end{align}
and the transmission energy consumption of user $m$ can be expressed  as
\begin{align}\label{eqemr}
E_m^R=P_m{t_m^R}.
\end{align}
When
task offloading is completed, the edge server computing process begins. We suppose the processing speed of
the
MR as $f_m^R$ in CPU cycles per second for
serving
user $m$. The
MR's
remote execution latency for user $m$ is given by
\begin{equation}\label{eqtre}
t_m^{eR} = \frac{{(1 - {\lambda _m}){d_m}{c_m}}}{{f_m^R}}.
\end{equation}
Since the size of
the remote execution
results
to the
users are extremely small, this part of time and energy costs can be neglected.
Similar to~(\ref{eqEl}),
the energy consumption of
the
MR to complete the task of
user
$m$ is
\begin{equation}\label{eqEr}
E_m^{{\rm{e}}R} = \xi {(f_m^R)^3}t_m^{eR} = \xi (1 - {\lambda _m}){d_m}{c_m}{(f_m^R)^2}.
\end{equation}

As we assume finite computation capacity at the MR,  a feasible computation resource allocation at the MR
should
satisfy $\sum\nolimits_{m = 1}^M {\sum\nolimits_{s = 1}^S {{x_{ms}}\textrm{y}_m^R} } f_m^R \le {f^R}$, where $f^R$ is the total computation  capacity of the MR in CPU cycles per second.

\smallskip
\subsubsection{BS Execution}

If the remote
execution location of
user $m$ is
chosen to be the BS,
we have ${\rm{y}}_m^B = 1$.
Since the offloaded data will still be forwarded by the MR,
we
call the transmission of computing tasks from user $m$ to MR and from MR to the BS as the first hop and the second hop, respectively.
It should be noted that, because the MR operates in the
FD mode,
the task transfers on the first hop and the second hop
are in parallel.
The MR receives data from
user $m$ while
forwarding
data to the BS at the same time
on the same
sub-channel.
The signal of the first hop will suffer from the  background noise as well as SI due to imperfect SI cancellation.
The specific cancellation methods are out of the scope of the
paper,
the value of
residual
SI after cancellation can be expressed in terms of the
transmit
power to facilitate the calculation. Specifically,  we use $\beta{P_m^R}$ to
represent
the residual
SI, where $\beta$ is
the SI cancelation level of the MR, which is a non-negative parameter, and $P_m^R$ is the
transmit
power of
the
MR
serving
user $m$.
The first hop SINR of user $m$ connected
to the MR through
sub-channel
$s$ can be expressed as
\begin{align}\label{eqSIN1}
\Gamma _{ms}^{B1} = \frac{{{H_s}{G_t}(m,R,s){G_r}(m,R,s)l_{mR}^{ - \alpha }{P_m}}}{{{N_0}W}+\beta{P_m^R}}.
\end{align}
The numerator of $\Gamma _{ms}^{B1}$ is exactly the same as the numerator of~\eqref{eqsinr},
except that we have
the residual
SI in the denominator.
The achievable
transmission rate of the first hop according to the Shannon formula
is
\begin{align}
  R_{ms}^{B1}=W\log_2(1+\Gamma _{ms}^{B1}).
\end{align}

Similar to~\eqref{eqsinr},
the SINR of the second hop from MR to the BS can be presented as
\begin{align}\label{eqSIN2}
\Gamma _{ms}^{B2} = \frac{{{H_s}{G_t}(R,B,s){G_r}(R,B,s)l_{RB}^{ - \alpha }{P_m^R}}}{{{N_0}W}},
\end{align}
where ${G_t}(R,B,s)$ and ${G_r}(R,B,s)$ are the transmitting gain and receiving gain of
the
directional antenna pointing from
the
MR to
the
BS on sub-channel $s$, respectively; $l_{RB}^{ - \alpha }$ is the path loss of
the MR-BS link with path loss exponent
$\alpha$; and
$P_m^R$ is the
transmit
power of
the
MR assigned to user $m$.

Similarly, the data rate of the second hop from the
MR to the
BS can be computed as
\begin{align}\label{eqRb2}
  R_{ms}^{B2}=W\log_2(1+\Gamma _{ms}^{B2}).
\end{align}
According to~\cite{FD-R}, the available data rate depends on the smaller
of
the first and second hop data rates
as
\begin{align}
R_{ms}^B = \min \{ R_{ms}^{B1},R_{ms}^{B2}\}.
\end{align}
Thus, the transmission delay consumed to upload $(1-\lambda_m)d_m$ data bits to
the
BS can be written as
\begin{align}\label{eqtB}
t_m^B = \frac{{(1 - {\lambda _m}){d_m}}}{{R_{ms}^B}}.
\end{align}
And the transmission energy consumption of user $m$ can be expressed  as
\begin{align}
E_m^B=P_m{t_m^B}.
\end{align}
Whereas,  the transmission energy consumption of
the
MR can be expressed in terms of
the transmit
power of
the
MR assigned to user $m$,
as well as
the
transmission time as
\begin{align}\label{eqEb}
E_m^{tR}=P_m^R{t_m^B} .
\end{align}

Denote $f_m^B$ as the processing speed of the BS in CPU cycles per second for
serving
user $m$. The BS's
remote execution latency for user $m$ is given by
\begin{align}
t_m^{eB} = \frac{{(1 - {\lambda _m}){d_m}{c_m}}}{{f_m^B}}.
\end{align}
Similarly, we also assume the total computing capacity of
the
BS is limited by $f^B$ in CPU cycles per second, as
$\sum\nolimits_{m = 1}^M {\sum\nolimits_{s = 1}^S {{x_{ms}}\textrm{y}_m^B} } f_m^B \le {f^B}$, where $f^B$ is the total computation  capacity of the BS in CPU cycles per second.

Hitherto we have
described the train-ground communication
mmWave
based MEC
system
and presented
the partial offloading model consisting of three cases:  local execution, MR execution, and BS execution. We proceed to formulate the joint user association, data segmentation, and resource allocation problem in
the
next section with the goal of
minimizing all the users' latency.

\section{Problem Formulation}\label{S4}

We aim to minimize the average task execution latency
of
all users,
which can partially offload their computation
tasks
to the BS or the MRs.
Once all users have determined the amount of data to
be offloaded
and the
target execution locations,
the portion of the data for remote execution is transferred over the
wireless
link to the associated BS or MR. When the
transmission
is completed,
the task
will be executed
by the remote server
at
the corresponding location. For partial
offloading,
there are two processes involved, namely local computation and
task offloading
(i.e.,
uploading task data
and
remote execution).
Since local computation can
be executed in parallel
with the computation offloading process, the total task computation delay for a user $m$ is determined by the longer process, given by
\begin{equation}\label{eqtm}
{t_m} = \max \{ t_m^L,{x_{ms}}\textrm{y}_m^R(t_m^R + t_m^{eR}),{x_{ms}}\textrm{y}_m^B(t_m^B + t_m^{eB})\}.
\end{equation}

With the system model in Section~\ref{S3}, we formulate a joint partial offloading, communication and computation resource allocation problem as follows.
\begin{subequations}
\begin{align}\label{eqa}
\hspace{-0.8in} \textbf{P1}: \ \min\limits_{{\bf{\lambda ,x,Y}},{\bf{f}},{{\bf{P}}^R}} \;\bar{t} = \frac{1}{M}{\sum\limits_{m = 1}^M {{t_m}} },
\forall m \in M 
\end{align}
\begin{align}\label{eqc1}
\mbox{s.t.}~~&0 \le {\lambda _m} \le 1 \\\label{eqc2}
&{\rm{y}}_m^R,{\rm{y}}_m^B \in \{ 0,1\} ,\forall m \in\textbf{ M} \\\label{eqc3}
&{\rm{y}}_m^R{\rm{ = 1}}\;{\rm{or}}\;{\rm{y}}_m^B = 1\;,if\;{\lambda _m} \ne 1 \\\label{eqc4}
&{\rm{y}}_m^R + \;{\rm{y}}_m^B \le 1 \\\label{eqc5}
&x_{ms}\in \{ 0,1\} ,\forall m \in\textbf{ M}, s\in \textbf{S} \\\label{eqc6}
&\sum\limits_{s = 1}^S {{x_{ms}}}  \le 1,\forall m \in \textbf{M} \\\label{eqc7}
&\sum\limits_{m = 1}^M {\sum\limits_{s = 1}^S {{x_{ms}}} } \le S \\\label{eqc8}
&\textrm{y}_m^R\left( {E_m^L + E_m^R} \right) \le {E_m},\forall m \in \textbf{M} \\\label{eqc9}
&\textrm{y}_m^B\left( {E_m^L + E_m^B} \right) \le {E_{\rm{m}}},\forall m \in \textbf{M} \\\label{eqc10}
&0 \le f_m^L \le f_m^{\max } \\\label{eqc11}
&\sum\limits_{m = 1}^M {\sum\limits_{s = 1}^S {{x_{ms}}\textrm{y}_m^R} } f_m^R \le {f^R} \\\label{eqc12}
&\sum\limits_{m = 1}^M {\sum\limits_{s = 1}^S {{x_{ms}}\textrm{y}_m^B} } f_m^B \le {f^B} \\\label{eqc13}
&\sum\limits_{m = 1}^M {\sum\limits_{s = 1}^S {{x_{ms}}} } \left( {\textrm{y}_m^RE_m^{eR} + \textrm{y}_m^BE_m^{tR}} \right) \le {E^R}.
\end{align}
\end{subequations}
%
%
Constraint~(\ref{eqc1})
specify the range of the data portion
that can be
executed
locally when user $m$ performs a partial
offloading.
The constraints relating to
user association are presented in~(\ref{eqc2})-(\ref{eqc4}).
Specifically,
(\ref{eqc3}) 
explains
that a certain number of tasks will be offloaded when the user establishes an association with the BS or an
MR,
whereas~(\ref{eqc4})
indicates
the user chooses
one of the MRs
or
the
BS as the remote execution location.
Constraints~(\ref{eqc5})-(\ref{eqc7})
are based on the fact that spectrum resources are limited.
Specifically,~(\ref{eqc6})
ensures
that a user can be assigned with
at most one sub-channel of the MR.
Constraint~(\ref{eqc7})
indicates that the total
number of
sub-channels assigned to users are limited by the total
number of available sub-channels.
Constraints~(\ref{eqc8}) and~(\ref{eqc9})
ensure that the energy consumption of
the
computation offloading process
of
user $m$, which consists of local computing and transmission energy consumption,
cannot
exceed the local energy budget $E_m$.
Constraint~(\ref{eqc10})
is on
the local CPU processing capability budget.
Constraints~(\ref{eqc11}) and~(\ref{eqc12})
ensures
feasible computation resource allocation at the MR and the BS, respectively. Finally,
constraint~(\ref{eqc13})
indicate
that the
execution
energy consumption and transmission energy consumption consumed
by an
MR are limited by $E^R$(i.e. The total energy of MR).

There is no doubt that
the formulated average latency minimization problem
is a mixed-integer nonlinear and non-convex programming problem.
Furthermore, the coupling of integer and continuous variables results in nonlinear constraints and non-convex feasible region.
In terms of complexity, the formulated problem is NP-hard, and is hard to solve
in polynomial time. Therefore, we
shall
propose an efficient and
practical
solution
algorithm
in the next sections.

\section{Problem Decomposition}\label{S5}

It can be seen that if the user association is determined,
constraints~(\ref{eqc1}) and~(\ref{eqc8})-(\ref{eqc10})
can be decoupled from the communication and computation resource allocation constraints~(\ref{eqc7}) and~(\ref{eqc11})-(\ref{eqc12}).
This implies that the problem can be solved by decomposition. For known user association, we first decide the data segmentation policy
considering
the energy and local CPU processing capability constraints~(\ref{eqc8})-(\ref{eqc10}).
Then, the original problem is transformed into
a
resource allocation problem under the energy consumption constraint of MRs
and computation resource
constraints of
MRs and the
BS.

\subsection{Data Segmentation Policy}\label{S4-2}

An important
question
for partial offloading is
how
to determine the optimal
partition
of data offloaded by a user, as it effects both the time consumption for local execution, offloading and remote execution, and the energy consumption for local computing and offloading. Based on
constraint~(\ref{eqc8})-(\ref{eqc10}),
an
energy efficient and local computation frequency bounded data segmentation policy can be derived.

Observing the objective function~(\ref{eqa}) and~(\ref{eqtm}),
we can see that the minimum latency for a user $m$ is reached when the two parallel processes (i.e.,
local execution and offloading plus remote execution) take the same
amount of
time. However, it is uncertain whether the user can equalize the time consumption of
these
two processes under various constraints.
We first examine
the upper bound and lower bound of the offloading fraction.
Assuming
that the association relationship of user $m$ has been established, then the original problem $\textbf{P1}$ can be transformed into the following problem.
\begin{align}\label{eqP2}
\textbf{P2}: & \min\limits_{\lambda_m,f_m^L}~\bar{t} = \frac{1}{M}{\sum\limits_{m = 1}^M {{t_m}} } \\
\mbox{s.t.}~&~\textrm{Constraints}~(\ref{eqc1}),~(\ref{eqc8})\sim(\ref{eqc10}). \nonumber
\end{align}
If $\textrm{y}_m^R =1$,
i.e.,
user $m$ is associated with an MR,
substituting~(\ref{eqEl}) and~(\ref{eqemr})
into the offloading energy consumption constraints~(\ref{eqc8}),
we obtain an inequality
related to $\lambda_m$ as
\begin{align}\label{eq4-2}
{\mu {\lambda _m}{d_m}{c_m}{{(f_m^L)}^2} + {P_m}\frac{{(1 - {\lambda _m}){d_m}}}{{R_{ms}^R}}} \le {E_m}.
\end{align}
Constraint~(\ref{eq4-2})
serves as the
feasibility condition in terms of
an
upper bound on $\lambda_m$, which can be easily obtained as
\begin{align}\label{eq4up}
{\lambda_m} \le \frac{{{E_m}R_{ms}^R - {P_m}{d_m}}}{{\mu {d_m}{c_m}{{(f_m^L)}^2}R_{ms}^R - {P_m}{d_m}}}.
\end{align}
As stated before, $\lambda_m$ is the
fraction
of data for
local computation of user $m$,
which
satisfies $0 \le {\lambda _m} \le 1$.
We
next obtain
an exact upper bound of $\lambda _m$ as
\begin{align}\label{eq4up1}
\lambda_m^{max1}=\min\left\{1, \ \frac{{{E_m}R_{ms}^R - {P_m}{d_m}}}{{\mu {d_m}{c_m}{{(f_m^L)}^2}R_{ms}^R - {P_m}{d_m}}}\right\}.
\end{align}
The above analysis also applies to the case where user
$m$
is associated with
the
BS (i.e., when
$\textrm{y}_m^B=1$). Similarly, substituting~(\ref{eqEl}) and~(\ref{eqEb})
into
constraints~(\ref{eqc9}), the
exact
supper bound of $\lambda _m$ in this case can be
obtained as
\begin{align}\label{eq4up2}
\lambda_m^{max2}=\min \left\{ 1, \ \frac{{{E_m}R_{ms}^B - {P_m}{d_m}}}{{\mu {d_m}{c_m}{{(f_m^L)}^2}R_{ms}^B - {P_m}{d_m}}}\right\}.
\end{align}

Based on the above analysis, we
obtain
the feasible set of $\lambda_m$,
which contains
the optimal solution of problem $\textbf{P2}$. First, we rewrite~(\ref{eqP2}) as
\begin{align}
\min\limits_{\lambda_m,f_m^L}{t_m}\rightarrow\min\limits_{f_m^L}\min\limits_{\lambda_m}{t_m}.
\end{align}
There are two independent variables in the
objective function
of problem $\textbf{P2}$. We first take one of the variables $f_m^L$ as a constant value, and then $t_m$ becomes a function
of $\lambda_m$.
Thus we can easily obtain the optimal solution $\lambda_m$, which minimizes $t_m$.
Secondly, we substitute the optimal $\lambda_m$ into $t_m$, where $t_m$ only depends on $f_m^L$, and find the value that minimizes $t_m$ under
constraint~(\ref{eqc10}).

We still focus on the case where
the user is associated with
an
MR as an example,
When
user $m$
is
associated with
an
MR, ${t_m}$ can be written as $\max \{ t_m^L,t_m^R + t_m^{eR}\}$.
Observing~(\ref{eqtl}),~(\ref{eqtr}), and~(\ref{eqtre}),
for a fixed $f_m^L$, $t_m^L$ is
a
monotonically increasing function of $\lambda_m$, while $t_m^R + t_m^{eR}$ is monotonically decreasing
with
$\lambda_m$.
Thus the $\lambda_m$ that
minimizes
${t_m}$
shall be
obtained when $t_m^L=t_m^R + t_m^{eR}$.
\begin{equation}\label{star}
\lambda_m^*=\frac{{f_m^L(f_m^R + {c_m}R_{ms}^R)}}{{f_m^L(f_m^R + {c_m}R_{ms}^R) + {c_m}R_{ms}^Rf_m^R}}.
\end{equation}
Here $\lambda_m^* $ is the value that makes the time of local execution and the time of MR
execution equal.
It's apparent from~(\ref{star})
that $0 \le \lambda _m^* \le 1$. Combining~(\ref{star}) and~(\ref{eq4up1})
where the possible values of $\lambda _m$ is given, we obtain
the optimal $\lambda_m$ as
\begin{align}\label{eqRopt}
\;\lambda _m^{R} = \left\{ {\begin{array}{ll} 
\lambda _m^*, & \mbox{if}~\lambda _m^* \le \lambda _m^{max1} \\
\lambda _m^{max1}, & \mbox{if}~\lambda _m^* > \lambda _m^{max1}.
\end{array}} \right.
\end{align}
Obviously, when $\lambda_m^* \le \lambda_m^{max1}$, we
choose the optimal $\lambda_m$ to be
$\lambda_m^*$. However, if $\lambda_m^* > \lambda _m^{max1}$,
$\lambda_m^* $ is out of
the feasible range of
$\lambda_m$.
As stated before, with
increased
$\lambda _m$, $t_m^L$ will increase while $t_m^R + t_m^{eR}$  will decrease.
We should
make these
two values as close as possible to minimize $t_m$. Consequently, the upper bound
$\lambda_m^{max1}$ will be taken for $\lambda_m$.

Similarly, we can obtain the optimal $\lambda_m$
when
user $m$
is associated
with the BS as follows.
\begin{align}\label{eqBopt}
\;\lambda _m^{B} = \left\{ {\begin{array}{ll} 
{\lambda'}_m^*, & \mbox{if}~{\lambda'}_m^* \le \lambda_m^{max2} \\
\lambda _m^{max2}, & \mbox{if}~{\lambda'}_m^* > \lambda_m^{max2},
\end{array}} \right.
\end{align}
where ${\lambda'}_m^*$ is given by
\begin{equation}\label{star'}
{\lambda'}_m^*=\frac{{f_m^L(f_m^B + {c_m}R_{ms}^B)}}{{f_m^L(f_m^B + {c_m}R_{ms}^B) + {c_m}R_{ms}^Bf_m^B}}.
\end{equation}

Note that the optimal $\lambda_m$s
in~(\ref{eqRopt}) and~(\ref{eqBopt})
still contain the unknown $f_m^L$.
For given
optimal $\lambda_m$,
the higher the local computing frequency $f_m^L$, the smaller the $t_m^L$, while the change in $f_m^L$ does not affect
the time it takes to complete the offloaded portion of the task.
Accordingly, to some extent,
a
higher $f_m^L$ can reduce $t_m$, and we take the optimal value of $f_m^L$ to be the largest value in~(\ref{eqc10})
as $f_m^{L}=f_m^{max}$.
In addition, considering that
some user's task is only executed locally,
if the user's energy consumption constraint cannot support the user to complete the local computing tasks at $f_m^{max}$, the frequency needs to be reduced according to~(\ref{eqEl}).
Hence, the optimal value of $f_m^L$ is given by
\begin{equation}\label{eqofL}
\;f_m^{*L}= \left\{ \begin{array}{ll} 
f_m^{max}, & \mbox{if}~f_m^{max} \le \sqrt {\frac{E_m}{\mu\lambda_m{d_m}{c_m}}} \\
\sqrt {\frac{E_m}{\mu\lambda_m{d_m}{c_m}}}, & \mbox{if}~f_m^{max} > \sqrt {\frac{E_m}{\mu\lambda_m{d_m}{c_m}}}.
\end{array} \right.
\end{equation}

\subsection{Power Allocation of MR}

When a user
offloads its tasks to the
BS, due to the
self-interference, the
transmit
power of the second hop link allocated
to the
MR
will cause interference to
the reception of
the first hop link, which will greatly affect the transmission rate of the
offloaded
task and further affect the user's delay. The power allocation scheme
for the
MRs is developed
in this section.

According to the SINRs and transmission rates
of the first and second hop links,
as well as the fact that the actual transmission rate from the user to BS depends on the smaller
one of the first and second hop links, we know that
when the value of $P_m^R$ equalizes the rates
of the
two hops, it can minimize the time
consumption for offloading and remote execution.
The optimal $P_m^R$ for user $m$ allocated by
the
MR is reached when the transmission rate of
the
two hops take the same value and given by
\begin{align}
P_m^{{\rm{opt}}R} = \frac{{ - {N_0}W}}{{2\beta }}{\rm{ + }}\frac{{\sqrt {{{({N_0}W)}^2}{b^2} + 4\beta {N_0}Wab} }}{{2\beta b}},
\end{align}
where
$a={H_s}{G_t}(m,R,s){G_r}(m,R,s)l_{mR}^{ - \alpha }{P_m}$ and $b={H_s}{G_t'}(m,R,s){G_r}(m,B,s)l_{RB}^{ - \alpha }$.

\subsection{Communication and Computation Resource Allocation}

With a given
user association and data segmentation strategy, we can transform the original problem {\bf P1}
as follows
for solving
the
communication and computing resource allocation problem, as
\begin{align}\label{eqP3}
\textbf{P3}: & \min\limits_{\textbf{Y}_m,x_{ms}, f_m^R,f_m^B}~\bar{t} = \frac{1}{M}{\sum\limits_{m = 1}^M {{t_m}} }\\
\mbox{s.t.}~&~~\textrm{Constraints}~(\ref{eqc2})\sim(\ref{eqc7}),~(\ref{eqc11})\sim(\ref{eqc13}). \nonumber
\end{align}
Problem $\textbf{P3}$ is still non-convex due to the product of integer and real valued variables. We simplify the computation resource  allocation constraint for remote execution as follows.
For the users offloading tasks to the MR, we adopt
the uniform resource allocation and obtain the computation resource allocated to a user at
the
MR as $f_m^R = f_R/|\mathcal {U}^R|$. Similarly, each user associated with BS obtain the computing resources as $f_m^B = f_B/|\mathcal {U}^B|$.

Solving the resource allocation problem will determine whether to associate with the BS or an MR, and how the sub-channels are accessed.
We
denote
the available
resources as
$\textbf{A}=\{k|k=(\textbf{Y}_m,s), s \in \textbf{S}\}$, with $\left| \textbf{A} \right| = 2 \times S$.
The
dual
selection of users and resources
can be regarded as
a matching problem,
i.e., modeled as a two-sided matching game.
To maximize their own benefits, the users in set $\textbf{M}$ are matched indepdently and rationally to the resources in set $\textbf{A}$.
Assume that the MR has perfect knowledge of the channel state information (CSI) of all users,
and
makes communication resources allocation decisions
using such information.
If $x_{ms}=1$ and $\textbf{Y}_m=(1,0)$ or $(0,1)$,
we have $x_{mk} = 1$ and
user $m$  and communication resource $k$ are matched with each other (i.e., a
matched
pair $(m, k)$
is formed).

\begin{definition}
Given two sets $\textbf{M}=\{1,2,...,M\}$ and
$\textbf{A}=\{k|k=(\textbf{Y}_m,s), s \in \textbf{S}\}$,
the user-resource matching  state $\Phi$ is a mapping from $m\in \textbf{M}$ to $k\in \textbf{A}$ and from $k\in \textbf{A}$ to $m\in \textbf{M}$. That is to say, it holds all established matching  pairs. Following are
the
details.
\begin{enumerate}
  \item $\Phi(m)\in \textbf{A},\forall m \in {\textbf{M}}$
  \item $\Phi (k) \in \textbf{M},\forall k \in {\textbf{A}}$
  \item $\left| {\Phi (m)} \right| \le 1$
  \item $\left| {\Phi (k)} \right| \le 1$
  \item $k \in \Phi (m) \Leftrightarrow m \in \Phi (k)$.
\end{enumerate}
\end{definition}
The above
definition
shows that the relationship between users and resources is one-to-one if
partial offloading
is performed.
In the case when all the tasks are executed locally, the sub-channels will be idle and no matched pair is found.
%
The criteria for establishing matched
pairs are based on the mutual
preferences
of both users and
resources.
Users are more inclined to choose resources that make their own parallel computing delay $t_m$ smaller.
Each user maintains a preference list of resources
in the
descending order.

\begin{definition}
For $k,k' \in \textbf{A},k \ne k'$, if $\Phi (m) = k,\Phi '(m) = k'$ (i.e., $\Phi$ and $\Phi'$ are two matched pairs),
we have
\[(k,\Phi ){ \prec_m}(k',\Phi ') \Leftrightarrow {t_m}(k,\Phi ) < {t_m}(k',\Phi ').\]
\end{definition}

A resource
element
prefers to select
the user that can
contribute to
minimizing
the total delay of
all
served users (including
both the
MR users and BS users). The preference order for resource $k\in \textbf{A}$ can be  defined as follows.

\begin{definition}
For $m,m' \in \textbf{M},m \ne m'$, if $\Phi (k) = m$, $\Phi '(k) = m'$, and $k=(\textbf{Y}_m,s )$, we have
\[(m,\Phi ){ \prec _k}(m',\Phi ') \Leftrightarrow \sum\limits_{s=1}^S {{x_{ms}}{t_m}(\Phi )}  < \sum\limits_{s=1}^S {{x_{ms}}{t_m}(\Phi ')}. \]
\end{definition}

The change of
the matched
pair of one user will have an effect on the total delay of all served users.
Under the energy consumption constraints of the MR, we define
the concept of swap-matching, which
help to further reduce
the latency of all users,
as well as
the concept of swap-blocking pair in the following.

\begin{definition}
Given a matching $\Phi$ with $\Phi(m)=k$, $\Phi(k) = m$, $\Phi(m')=k'$, and $\Phi(k') = m'$, if a swap-matching occurs,  that is, $\Phi _{mk}^{m'k'}{\rm{ = }}\Phi \backslash \{ (m,k),(m',k')\}  \cup \{ (m,k'),(m',k)\}$,
the matching pairs
should be updated to
$\Phi(m)=k',\Phi(k') = m,\Phi(m')=k,\Phi(k) = m'$.
\end{definition}

\begin{definition}
For a pair $(m,m')$ with $\Phi(m)=k$, $\Phi(k)=m$, $\Phi(m')=k'$, and $\Phi(k')=m'$,
if the following conditions are satisfied
\begin{enumerate}
  \item $(m,\Phi _{mk}^{m'k'}){ \prec _{k'}}(m',\Phi)$,
  \item $(m',\Phi _{mk}^{m'k'}){ \prec _k}(m,\Phi)$
  \item $(k,\Phi _{mk}^{m'k'}){ \prec _{m'}}(k',\Phi)$
  \item $(k',\Phi _{mk}^{m'k'}){ \prec _m}(k,\Phi)$,
\end{enumerate}
then $(m,m')$ forms a swap-blocking pair, which means that if $m$ and $m'$ swap their matching resources with each other, both users and resources will be more satisfied. 
Taking 1) and 3) as an example, resource $k'$ prefers to match user $m$ rather than $m'$,
and user $m'$ prefers to match resource $k$ rather than $k'$.
\end{definition}

\section{The Proposed JRACO Scheme}\label{S6}

In this section, we
present
the proposed joint resource allocation and computational offloading
scheme
(JRACO)
with MR energy consumption constrained.
JRACO
mainly
comprises
two parts. The first part deals with the joint
problem of
communication and computing resource allocation and computing offloading
aiming to minimize the delay of all users, without considering the MR energy consumption constraints in \ref{eqc13}. The result of the first
part
is then used as input to the second
part,
which consists of
the relevant measures to control the energy consumption
under the MR energy
constraint (i.e. $E^R$).

\subsection{Resource Allocation and Computation Offloading Algorithm}

Due to limited communication resources,
we should first establish a criterion to
screen out those users who can
be served with computational offloading.
As described above, there is a certain connection between
an
MR link and
an
FD-BS link, and a sub-channel can only be assigned to one MR link or one FD-BS link.
In order to simplify the analysis of selecting service users, we assume that the users being served are first associated with
the
MR. After determining which users can be served, we then design an association scheme for each user.

\begin{algorithm}[!t] 
\caption{The Resource Allocation and Computation Offloading (RACO) Algorithm}
\label{alg1}
\begin{algorithmic}[1]
\begin{small}
\REQUIRE $\lambda_m=1,{\textbf{Y}_m}=(\textrm{y}_m^R,\textrm{y}_m^B) = (0,0),x_{ms}=0,\forall m\in \textbf{M},s\in \textbf{S},\Phi=\varnothing,num\_ac = 0$            
\ENSURE $\lambda_m,f_m^R,f_m^B,,x_{ms},\forall m\in M,\mathcal {U}^R,\mathcal {U}^B$
\STATE \textbf{S}'=\textbf{S}, $\textbf{M}'=$ User set in descending order of $d_m$ ;
\FOR {$ m\in \textbf{M}'$ }
\STATE Find $s=\textrm{argmax}_s R_{ms}^R$,$s\in \textbf{S}'$ ;
\IF {$M \leq S$}
\STATE ${x_{ms}} = 1,\textbf{Y}_m =(1,0),\textbf{S}'= \textbf{S}'\setminus s,\Phi  = \Phi  \cup (m,k)$;
\ELSE
    \IF{$num\_ac < S$}
        \IF {$\frac{{{d_m}{c_m}}}{{f_m^L}} > \frac{{{d_m}}}{{R_{ms}^R}}+ \frac{{{d_m}{c_m}}}{{f_m^R}}$, where $f_m^R = \frac{f^R}{S/2}$ }
            \STATE ${x_{ms}} = 1,\textbf{Y}_m =(1,0),\textbf{S}'=\textbf{S}'\setminus s,\Phi  = \Phi\cup (m,k),num\_ac=num\_ac+1$;
        \ENDIF
    \ELSE
        \STATE $s=\textrm{argmax}_s{R_{ms}^R},s\in \textbf{S}$, and find $m_1$ where $x_{{m_1}s}=1$;
        \IF {$\frac{{d_m}{c_m}}{f_m^L} - (\frac{d_m}{R_{ms}^R} + \frac{{d_m}{c_m}}{f_m^R}) > \frac{{d_{m_1}}{c_{m_1}}}{f_{m_1}^L} - (\frac{{{d_{{m_1}}}}}{{R_{ms}^R}} + \frac{{{d_{{m_1}}}{c_{{m_1}}}}}{{f_m^R}})$}
            \STATE ${x_{ms}} = 1,x_{{m_1}s}=0,\textbf{Y}_m =(1,0),\textbf{Y}_{m_1} =(0,0),\Phi  = \Phi\setminus (m_1,k)  \cup (m,k)$;
        \ENDIF
    \ENDIF
\ENDIF
\ENDFOR
\STATE All users accessing resources are recorded in set $\mathcal {U}^R$;
\REPEAT
\STATE Calculate ${t_m}(\lambda _m^{optR},f_m^{opt})$ for $m\in \mathcal {U}^R$;
\FOR{$m\in \mathcal {U}^R$}
    \FOR{$m'\in \mathcal {U}^R$}
    \IF {$m\neq m'$ and $(m,m')$ is swap-blocking pair with $\Phi(m)=k,\Phi(m')=k' \in\Phi$}
    \STATE $\Phi  \leftarrow \;\Phi _{mk}^{m'k'},{x_{ms'}} = {x_{m's}} = 1,{x_{ms}} = {x_{m's'}} = 0$;
    \ENDIF
    \ENDFOR
\ENDFOR
\UNTIL there is no swap-blocking pairs existed in $\Phi$
\STATE Find the best number of BS users $num_B$ with $\min\sum\limits_{m \in \mathcal {U}^R}{{t_m}(\lambda_m^{opt})},\lambda_m^{opt}\in(\lambda_m^{optR},\lambda_m^{optB})$;
\STATE Update $f_m^{optR}=\frac{f_R}{\left|\mathcal {U}^R\right|-num_B},f_m^{optB}=\frac{f_R}{num_B}$;
\FOR {$m\in \mathcal {U}^R$}
\STATE $\Delta t_m={t_m}(\lambda_m^{optR})-{t_m}(\lambda_m^{optB})$;
    \IF {$\Delta_m>0~ \&\&~ E_m^{eR}>E_m^{tR}$}
    \STATE $\mathcal {U}^R\setminus m, \mathcal {U}^B\cup m$;
    \ENDIF
\ENDFOR
\IF{$\left|\mathcal {U}^B\right|>num_B$}
\STATE Remove the user with the smallest $\Delta t_m$ in $\mathcal {U}^B$ to $\mathcal {U}^R$ until $\left|\mathcal {U}^B\right|=num_B$;
\ELSIF {$\left|\mathcal {U}^B\right|<num_B$}
\STATE Add $m' \in \mathcal {U}^R$ to $\mathcal {U}^B$ until $\left|\mathcal {U}^B\right|=num_B$, $m'$ has the smallest ${t_m}(\lambda_m^{optB})$ among $\mathcal {U}^R$ with $\Delta t_{m}\geq 0$;
\IF{$\left|\mathcal {U}^B\right|<num_B$}
\STATE Add $m''$ to $\mathcal {U}^B$ until $\left|\mathcal {U}^B\right|=num_B$, $m''$ has the biggest $\Delta t_{m}$ among $\mathcal {U}^R$ with $\Delta t_{m}<0$;
\ENDIF
\ENDIF
\end{small}
\end{algorithmic}
\end{algorithm}

The proposed RACO algorithm is presented in Algorithm~\ref{alg1}.
If there is no reasonable admission control,
the
initial parameter setting
could
be
infeasible.
we assume that all users are performing on the local platform in the initial state, i.e., $\lambda_m=1$, for all $m\in \textbf{M}$.
The parameter $num\_ac$ records the usage of resources. And the set $M'$ records the users in descending order of $d_m$. Users with heavy computing tasks have priority in choosing communication resources.
Steps 3-10
are for the case when there are sufficient sub-channels
and Steps 12-15
are for the case when the sub-channel resource
is
deficit.
If there is sufficient resources,
users
will choose the available sub-channel with the maximum transmission rate
as given
in Step 3. Specifically, if the number of users is less than the number of sub-channels, then the
users
are
directly matched
with
the sub-channels and MRs;
Otherwise, the degree of demand for resources will be evaluated.
If the  latency
of fully
edge computing is greater than that of
fully MR computing for user $m$,
sub-channel $s$
will be
occupied by
user
$m$ and the
matched
pair $(m,k)$ will be
approved.
Note that the CPU frequency of the MR here is
given by $2 f^R/S$.
Since the final number of MR users is not yet known, we assume that the
sub-channels
are equally
assigned
to MR users and BS users.
We also take into account
the case of
resource deficit. In
Step 12,
the
user still prefers the sub-channel
$s$
with the best transmission performance, but will compete with
user $m_1$ that previously obtained the sub-channel.
Considering the optimization objective, the user with
a
greater difference between the delay of
fully
remote execution
and that of
fully
local
execution
has
a higher privilege
in accessing
resource $s$,
The
criterion in
Step 13 not only requires user $m$
to satisfy
the sufficient and necessary conditions
for offloading
(see Step 8), but also has
a higher privilege
than the
previously matched
user $m_1$.
In this case, user
$m$ is accepted,
user
$m_1$ is rejected, and
we have
$\Phi  = \Phi\setminus (m_1,k)  \cup (m,k)$.

So far, the algorithm has determined
which users can be served and makes
the initial sub-channel allocation close to the optimization goal, thus reducing the complexity of
the
subsequent matching game. All the served users are temporarily associated with
the
MR. After determining the best user-sub-channel matching
through the matching game in
Steps 20-29,
the algorithm then decides for each user whether to be associated with the MR or the BS depending on its best sub-channel in Steps 30-45.

The set
$\mathcal {U}^R$ sorts out all users that need
to partially offload their tasks.
From
Steps
20-29, in one iteration,
the algorithm first obtains
$\lambda_m^{optR}$
and
$f_m^{opt}$  for
each
$m\in \mathcal {U}^R$. Based on these values, all user pairs are
examined
to find the
candidate
swap-blocking pair.
Then
swap matching is triggered, and resource allocation changes dynamically as the matching game evolves. The condition under which the alternating iteration is ended is that $\left|\mathcal {U}^R\right|\ast(\left|\mathcal {U}^R\right|-1)$ user pairs are not swap-blocking pairs.

In Step 30, 
the algorithm
changes
the number of users associated with
the
BS in $\mathcal {U}^R$, to find the optimal $num_B$ that minimizes the total delay.
After obtaining
the optimal $num_B$,
Steps 31-45 determine which users should be associated with the BS.
First,
according to the optimal $num_B$, the optimal BS and MR CPU
frequencies
for each user are obtained in Step 31.
Then
the latency of the respective
MR
association
and BS
association
for each user is derived,
whose difference is recorded at $\Delta t_m$. In Steps 32-37,
the algorithm
selecting
the
BS users
for
$\mathcal {U}^B$ according to the time delay and energy consumption.
These
two metrics are better for users in $\mathcal {U}^B$ than they would be if they were associated with MR. Note that, considering the energy consumption constraints of
the
MR, the energy
consumption
of
the
BS users is related to
wireless
transmission, while that of
the
MR users is related to
the
calculation in~\eqref{eqEr}. 
The number of BS users selected
in the above steps
is not necessarily equal to the optimal $num_B$.
If it is larger than the optimal $num_B$,
the algorithm
changes
the associations of those
users whose latency
will be
affected slightly by this change in Step 39.
Otherwise,
we need to add
some
users from $\mathcal {U}^R$ to $\mathcal {U}^B$.
Each time the user
is selected who
has
the lowest latency if associated with
the
BS and $\Delta t_m>0$, until the
optimal
$num_B$ is reached.
Here $\Delta t_m>0$ indicates that
it is better for
the
user
to choose
the
BS than
the
MR in terms of delay.
If all
the
users with $\Delta t_m>0$ have been selected but
we still have
$\left|\mathcal {U}^B\right|<num_B$, then
the
users with $\Delta t_m<0$
will be selected.
These users actually prefer to
be
associated with
the
MR.
Thus, we choose users with
a
small difference in delay between
these
two types of
associations,
while
the biggest $\Delta t_m$ indicates the smallest difference in delay in these two cases.

The
computational complexity
of Algorithm~\ref{alg1}
mainly
comes from three parts.
First, from Steps 2-18,
most computations are from Step 3
and
Step 12 for each user,
with complexity
$O(MS)$.
The second
part
comes from
Steps 20-29,
where each iteration
takes $O({\left|\mathcal {U}^R\right|}^2)$
to determine all
the
swap-blocking pairs. We assume the algorithm converges after $I$ iterations.
Then
the complexity is $O(I{\left|\mathcal {U}^R\right|}^2)$. Finally, in Steps 30-45, It takes $O(\log_2{\left|\mathcal {U}^R\right|})$ iterations to find
$num_B$
using
binary search in Step 30, and $O(\left|\mathcal {U}^R\right|)$
to determine the association. Thus,
the computational
complexity of the proposed algorithm is $O(MS+I{\left|\mathcal {U}^R\right|}^2)$ for $\left|\mathcal {U}^R\right|<M$.

\subsection{MR Energy Constraint Algorithm}

The proposed RACO algorithm does not consider the energy constraint of
the
MR.
On
the basis of the former algorithm, we need to
ensure that
the sum of the MR energy consumption of users
does not exceed $E^R$ in constraint (\ref{eqc13}).

This problem can be transformed into a
knapsack
problem.
The energy constraint of
the
MR
is regarded as
capacity of the
knapsack,
the MR energy consumed by each served user
given by
Algorithm~\ref{alg1}
is equivalent to the weight of each
item,
and the partial offloading delay ($t_m$) is regarded as the value of the item. The smaller the delay, the higher the value.
The problem to be solved
here
is, how to select the
items
to put in the
knapsack,
without violating the
knapsack's
capacity and maximize the total value.
Specifically, we model the problem as a decimal knapsack problem, where
a portion of an item can be selected and loaded
into the knapsack, because it's easy to accomplish by reducing the offloaded data bits. Tasks for users who cannot fit into
the
knapsack
will
be
executed locally.

\renewcommand{\algorithmicrequire}{\textbf{Input:}}
\begin{algorithm}[!t]
\caption{The MR Energy Constraint Algorithm}
\label{alg2}
\begin{algorithmic}[1]
\begin{small}
\REQUIRE $\mathcal {U}^R$ and $\mathcal {U}^B$
\ENSURE $\bar{t} = \frac{1}{M}{\sum\nolimits_{m = 1}^M {{t_m}}},m\in \textbf{M}$
\STATE $E=$ the total MR energy of all users in $\mathcal {U}^R$ and $\mathcal {U}^B$;
\IF {$E>E^R$}
\STATE $\textbf{U}=$ user set in ascending order of $\frac{t_m}{E_m^{*R}},E_m^{*R}\in \{E_m^{eR},E_m^{tR}\}$;
\FOR {$m\in \textbf{U}$}
  \STATE Update $\mathcal {U}^R,\mathcal {U}^B$ and $\mathcal {U}^\times$, where $E\leq E^R,E+E_k^{*R}>E^R$,
  and $k$ is the
  first
  user in $\mathcal {U}^\times$;
\ENDFOR
\IF{$E^R-E>0$}
\IF {$k$ is a MR user}
    \REPEAT
    \STATE Update $f_k^R=\sqrt{\frac{E^R-E}{\xi(1-\lambda_k^R){d_k}{c_k}}}$, $\lambda_k^R,\mathcal {U}^R,E_k^{eR}$;
    \UNTIL {$0\le E^R-(E+E_k^{eR})\leq\varepsilon_1$}
\ELSIF {$k$ is a BS user}
    \STATE Update $\lambda_k^B=1-\frac{(E^R-E)R_{ks}^B}{P_k^R{d_k}},t_k(\lambda_k^B),\mathcal {U}^B$;
\ENDIF
\STATE Calculate $T_1=\sum {t_m},m\in \textbf{M}$; 
\IF {$\mathcal {U}^\times$ contains MR users}
    \REPEAT
    \STATE $\mathcal {U}^\times\setminus m',\mathcal {U}^R\cup m',m'$ is the $1th$ user in $\mathcal {U}^\times$;
    \WHILE{$E^R-E<0$}
    \STATE $f_n^R=f_n^R-\varepsilon_2$, update $\lambda_{n}^R,t_{n},n\in\mathcal {U}^R$;
    \STATE Calculate $T_2=\sum {t_m},m\in \textbf{M}$;
    \ENDWHILE
    \UNTIL{$T_2$ is no longer decreasing or no MR user in $\mathcal {U}^\times$}
\ENDIF
\STATE
$\bar{t} = \frac{1}{M}\min \{T_1, T_2\}$;
\ENDIF
\ENDIF
\end{small}
\end{algorithmic}
\end{algorithm}

The proposed Algorithm~\ref{alg2} consists of two parts.
The first part is
the greedy algorithm
in
Steps 4-14, and the second part is
the heuristic algorithm
in Steps
16-24. Finally, our optimization goal,
minimizing
the average delay of all users,
will be achieved as
the smaller one
of the results of these two parts.

Specifically,
we should
first
determine
whether the total MR energy consumption of the served users
given by
Algorithm 1 exceeds $E^R$. If not, the result of Algorithm 1
will be
the final optimal solution, and there is no need for executing this algorithm anymore.
Otherwise,
the rest of the algorithm in Steps 3-26 will be executed.
In Step 3, the value rate of all service users was sorted from high to low and recorded in set $\textbf{U}$, where the lower the delay per unit of energy consumption, the higher the rate.
In
Steps 4-6, users are put into the
knapsack
in the order of $\textbf{U}$, until the remaining capacity cannot
accomodate
a user. Users that cannot
be put into the
knapsack
will be
put into $\mathcal {U}^\times$ temporarily.
Users who have been placed in the
knapsack
must be served. If there is still capacity left in the
knapsack,
the next steps in Lines 8-25 will be
executed
to reduce the delay of all users as much as possible.
Otherwise,
all the rejected
users
in $\mathcal {U}^\times$
will not
perform partial offloading and
their tasks will be executed locally.
In
Steps 8-14, the remaining
knapsack
capacity is first used to load the first user in $\mathcal {U}^\times$(i.e. user $k$), because
the user has the highest rate.
As mentioned above, only a
part of user $k$'s task
can be
fitted into the
knapsack,
which means that measures need to be taken to reduce the MR energy consumption
for this user
to $(E^R-E)$.
If user
$k$ is associated with
the
MR, we can keep
$E_k^{eR}$ within the constraint by reducing $f_k^R$ according to~(\ref{eqEr}).
As $f_k^R$
is decreased,
the
offloaded
fraction of
the
task will be adjusted accordingly; $f_k^R$ and $\lambda_k^R$ interact with each other
iteratively,
so that $E_k^{eR}$ converges to $E^{R}$.
If
user
$k$ is associated with
the
BS, we can keep the $E_k^{tR}$ within the constraint by reducing the
offloaded fraction (i.e. $1-\lambda_k^B$) according to~(\ref{eqtB}) and~(\ref{eqEb}).
Finally, all users in $\mathcal {U}^\times$ except $k$ are denied service, and all their tasks
will be executed
locally.

In
Steps 16-24, we use another strategy to constrain the MR energy consumption of served users.
According to~(\ref{eqEr}), decreasing $f_m^R$ linearly
will reduce
$E_m^eR$ exponentially, 
Thus,
we appropriately reduce $f_m^R$ of users in $\mathcal {U}^R$ obtained in
Step 5,
leaving more MR energy to serve
the
MR users in $\mathcal {U}^\times$.
That is to say, the latency of the
users
in $\mathcal {U}^R$
will be
sacrificed for
reducing
the total delay.
This is because
the computing capacity of
the
MR is far better than that of the local device.
In
Steps 17-23,
the number of served MR users is gradually increased,
and the $f_m^R$ of the users in the set
is
adjusted uniformly every time
$\mathcal {U}^R$ is updated.
When the total delay of users
cannot
be further
reduced or when
there is no MR user in $\mathcal {U}^\times$,
we
stop the iteration.
In
Steps
19-22,
$f_m^R$ is adjusted
by setting a reasonable step size, and
the optimal
offloaded
fraction of each MR user
is updated
according to~(\ref{eqRopt}),
until the total MR energy consumption of served users
becomes
lower than $E^R$.
Finally,
Step
25 indicates that the solution to problem \textbf{P1} is obtained by
comparing
the outcomes of the
two schemes in Algorithm~\ref{alg2}.

The computation complexity of Algorithm~\ref{alg2}
mainly lies in the iterative
Steps
9-11 and 17-23.
In Steps 17-23, the outer loop is executed at most $\left|\mathcal {U}^\times\right|$ times.
For the inner loop, the number of iterations of
the
{\em while} loop is determined by the accuracy tolerance $\varepsilon_2$. Given $\varepsilon_2>0$, the complexity of
the
one dimensional search on $f_m^R$ is
$O(\log(1 / \varepsilon_2))$.

\section{Performance Evaluation}\label{S7}


\subsection{Simulation Setup}

In this section, we evaluate the performance of the proposed schemes
by comparing
with several
baseline
schemes, and the simulations are conducted in MATLAB.
In the simulation scenario, Multiple users are scattered in
a
circular area
centered at the MR
with a radius of 120$m$,
and a BS is 500$m$ away from the MR.
Different local device applications can be distinguished by the size of computing data $d_m$, which
follows a
uniform distribution between [1,4]$Mbits$.
The computation workload $c_m$
follows
the uniform distribution between [300,500]$cycles/bit$.
The maximum local computation capacity $f_m^{max}$ follows the uniform distribution between [0.3,0.5]$GHz$.
The local energy constraint $E_m$ for each user is randomly chosen
between
$\{0.5,1.2,1.8\}J$.
unless otherwise specified, the MR and BS computation capacity $f^R$ and $f^B$ are set to
$8GHz$ and $24GHz$, respectively.
The
mmWave
frequency band
used in the simulations
is 28$GHz$ and the bandwidth is 2$GHz$. In addition, the channel
follows the
Nakagami distribution
with parameters $m_s=3$ and $w_s=\frac{1}{3}$ \cite{yali} and the mm-wave realistic directional antenna model from IEEE 802.15.3c \cite{80215}.
Other parameter settings are
provided in Table~\ref{table2}. The results of simulation studies in this
section are based on an average over a number of Monte
Carlo simulations for various system parameters.
The benchmark schemes
are as follows.


\begin{table}[!t]
\begin{center}
\caption{Simulation Parameters}
\begin{tabular}{lll}
\toprule
Parameter & Symbol  & Value \\
\midrule
Noise spectral density & $N_0$ & -134 dBm/MHz \\
User transmit
power & $P_m$ & 5 dBm \\
Path loss exponent  &  $\alpha$   &  3 \\
SI cancelation level & $\beta$ & $10^{-12}\sim10^{-11}$ \\
Half-power beamwidth  &   $\theta_{-3dB}$ &    $30^{\circ}$ \\
Effective switched capacitance & $\mu$ & $5 \times 10^{-27}$\\
Energy accuracy & $\varepsilon_1$ & $0.001$ J \\
Frequency accuracy & $\varepsilon_2$ & $0.002$ GHz \\
\bottomrule
\end{tabular}
\label{table2}
\end{center}
\end{table}

\begin{itemize}
\item \textbf{USRA}: When determining the served users,
the order of users accessing sub-channel resources is random.
The served users still determine the optimal
sub-channel
through the matching game, but it is random for users to associate with
the
MR or
the
BS on their optimal
sub-channel.
In addition, only the first part of Algorithm~\ref{alg2}
is used for
enforcing
the MR energy consumption constraint.

\item \textbf{RUNP}: Select the served users randomly according to the number of sub-channels. For those users associated with the BS, the
transmit
power
for the
MR, $P_m^R$,
is randomly selected from $[0.1,0.6]$ W.
The remaining parts
are the same as that in our scheme.

\item \textbf{RO}:
The fraction of task data to be offloaded
for
each served user is decided randomly, while user association and matching game are
the
same as
that in
our scheme.
Only the first part of Algorithm~\ref{alg2}
is used for
enforcing the
MR energy consumption constraint.

\item \textbf{JPORA}:
Similar to the existing work in \cite{D2D},
the user association problem is solved based on the location of users,
users within the BS coverage radius are associated with it.
Subcarrier assignment of a user based on maximum marginal data rate.
Optimizing offloading ratio iteratively, iteration ends when the average local computation delay and average computation offloading delay of all users are equal.
\end{itemize}
In Table \ref{CIS}, \ref{CIUE}, we give 95\% confidence interval with the normal distribution analysis of several random algorithms in Fig.~\ref{fig1} and Fig.~\ref{fig2}, where mu is the mean and sigma is the standard deviation.

\subsection{Comparison with Baseline Schemes}

\begin{figure}[!t]
\begin{center}
\includegraphics*[width=1\columnwidth,height=2.5in]{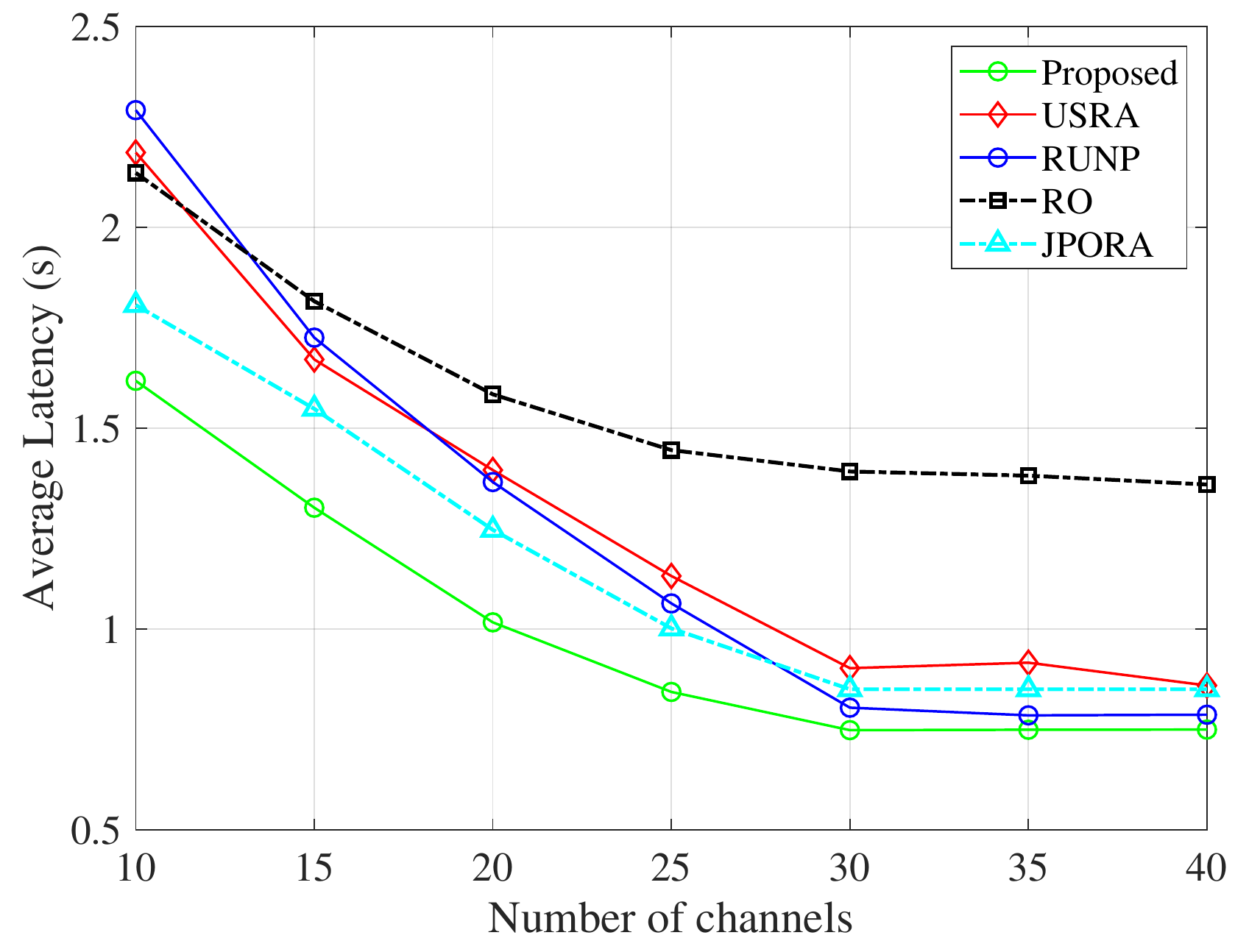}
\end{center}
\caption{Average latency of
the
four schemes with different
numbers
of sub-channels.}
\label{fig1}
\end{figure}

\begin{table*}
  \centering
  \fontsize{9}{12}\selectfont
  \caption{CONFIDENCE INTERVAL ANALYSIS OF DIFFERENT NUMBER OF CHANNELS}
  \label{CIS}
    \begin{tabular}{|c|c|c|c|c|c|c|}
    \hline
    \multirow{2}{*}{Number of channels}& \multicolumn{2}{c|}{USRA}&\multicolumn{2}{c|}{RUNP}&\multicolumn{2}{c|}{RO}\cr\cline{2-7}
    & mu & sigma & mu & sigma & mu & sigma\cr
    \hline
    \multirow{7}{*}{}
     $10$&[2.1667,2.2109]&[0.0976,0.1292]&[2.2829,2.3042]&[0.0472,0.0625]&[2.1237,2.1496]&[0.0572,0.0756]\cr\cline{1-7}
     $15$&[1.6467,1.6878]&[0.0910,0.1203]&[1.7180,1.7397]&[0.0482,0.0637]&[1.8005,1.8242]&[0.0523,0.0692]\cr\cline{1-7}
     $20$&[1.3720,1.4153]&[0.0958,0.1268]&[1.3469,1.3688]&[0.0484,0.0641]&[1.5662,1.5889]&[0.0503,0.0665]\cr\cline{1-7}
     $25$&[1.1864,1.2323]&[0.1015,0.1343]&[1.0825,1.1055]&[0.0509,0.0673]&[1.4270,1.4509]&[0.0529,0.0700]\cr\cline{1-7}
     $30$&[0.8895,0.9307]&[0.0912,0.1206]&[0.7871,0.7940]&[0.0150,0.0198]&[1.3717,1.3986]&[0.0596,0.0788]\cr\cline{1-7}
     $35$&[0.8683,0.9013]&[0.0729,0.0964]&[0.7839,0.7873]&[0.0075,0.0099]&[1.3669,1.3894]&[0.0496,0.0656]\cr\cline{1-7}
     $40$&[0.8732,0.9098]&[0.0811,0.1073]&[0.7835,0.7867]&[0.0072,0.0095]&[1.3416,1.3648]&[0.0513,0.0679]\cr\cline{1-7}
    \hline
    \end{tabular}
\end{table*}

\begin{table*}
  \centering
  \fontsize{9}{12}\selectfont
  \caption{CONFIDENCE INTERVAL ANALYSIS OF DIFFERENT NUMBER OF USERS}
  \label{CIUE}
    \begin{tabular}{|c|c|c|c|c|c|c|}
    \hline
    \multirow{2}{*}{Number of users}& \multicolumn{2}{c|}{USRA}&\multicolumn{2}{c|}{RUNP}&\multicolumn{2}{c|}{RO}\cr\cline{2-7}
    & mu & sigma & mu & sigma & mu & sigma\cr
    \hline
    \multirow{7}{*}{}
     $15$&[0.6121,0.6493]&[0.0823,0.1089]&[0.3612,0.3666]&[0.0120,0.0159]&[1.2129,1.2358]&[0.0506,0.0669]\cr\cline{1-7}
     $20$&[0.6707,0.7060]&[0.0780,0.1032]&[0.3619,0.3677]&[0.0129,0.0170]&[1.2503,1.2734]&[0.0510,0.0675]\cr\cline{1-7}
     $25$&[0.6470,0.6833]&[0.0804,0.1064]&[0.4195,0.4271]&[0.0168,0.0223]&[1.2461,1.2680]&[0.0464,0.0649]\cr\cline{1-7}
     $30$&[0.7187,0.7445]&[0.0770,0.0954]&[0.5243,0.5427]&[0.0407,0.0538]&[1.3873,1.4102]&[0.0507,0.0671]\cr\cline{1-7}
     $35$&[1.0329,1.0659]&[0.0729,0.0964]&[0.7373,0.7577]&[0.0450,0.0596]&[1.4719,1.4953]&[0.0518,0.0685]\cr\cline{1-7}
     $40$&[1.2343,1.2653]&[0.0685,0.0907]&[0.9881,1.0106]&[0.0499,0.0660]&[1.6014,1.6231]&[0.0481,0.0637]\cr\cline{1-7}
     $45$&[1.6042,1.6372]&[0.0730,0.0965]&[1.3529,1.3750]&[0.0489,0.0647]&[1.9280,1.9511]&[0.0511,0.0677]\cr\cline{1-7}
    \hline
    \end{tabular}
\end{table*}

As
communication resource allocation is a challenging part of
the
problem, we need to determine the available sub-channels to verify the application of our scheme in practical scenarios.
In Fig.~\ref{fig1}, the numbers of users are set to
30. When resources are insufficient, that is, the number of sub-channels varies from $10$ to $25$, all
the
four schemes enable more users to take advantage of
the
parallelism by partitioning
and offloading
part of
the
tasks to edge computing.
The average latency is observed to decrease
with increased number of channels.
The similar performance of USRA and RUNP shows that sometimes communication resource matching is superior to computation resource allocation, and sometimes vice versa.
Although RO has the worst performance in most cases, it can even outperform USRA and RUNP when communication resources are scarce, with a well-chosen set of users.
When the number of sub-channels exceeds 25, all users have
the
opportunity to offload
their tasks,
and the average latency tends to be roughly constant except for USRA. Because the random association may
cause
large amounts of users
to be
associated with
the
BS or
the
MR,
and the resource allocated
to each user is small, which makes the latency larger.
Random assignment of some computing tasks to the edge server is
clearly helpful
to reduce the performance of RO compared with the proposed algorithm.
Compared with RUNP, it is verified that the served user selection scheme in the proposed algorithm can
effectively
reduce
the overall delay
of all users.
JPORA adjusts the offloading ratio uniformly for all users, resulting some users didn't get the optimal offloading ratio when the algorithm is stopped.
According to the confidence interval, among the three random algorithms, RUNP is more stable when resources are surplus. While USRA has the highest uncertainty, indicating the importance of reasonable allocation of computation resources.

\begin{figure}[!t]
\begin{center}
\includegraphics*[width=1\columnwidth,height=2.5in]{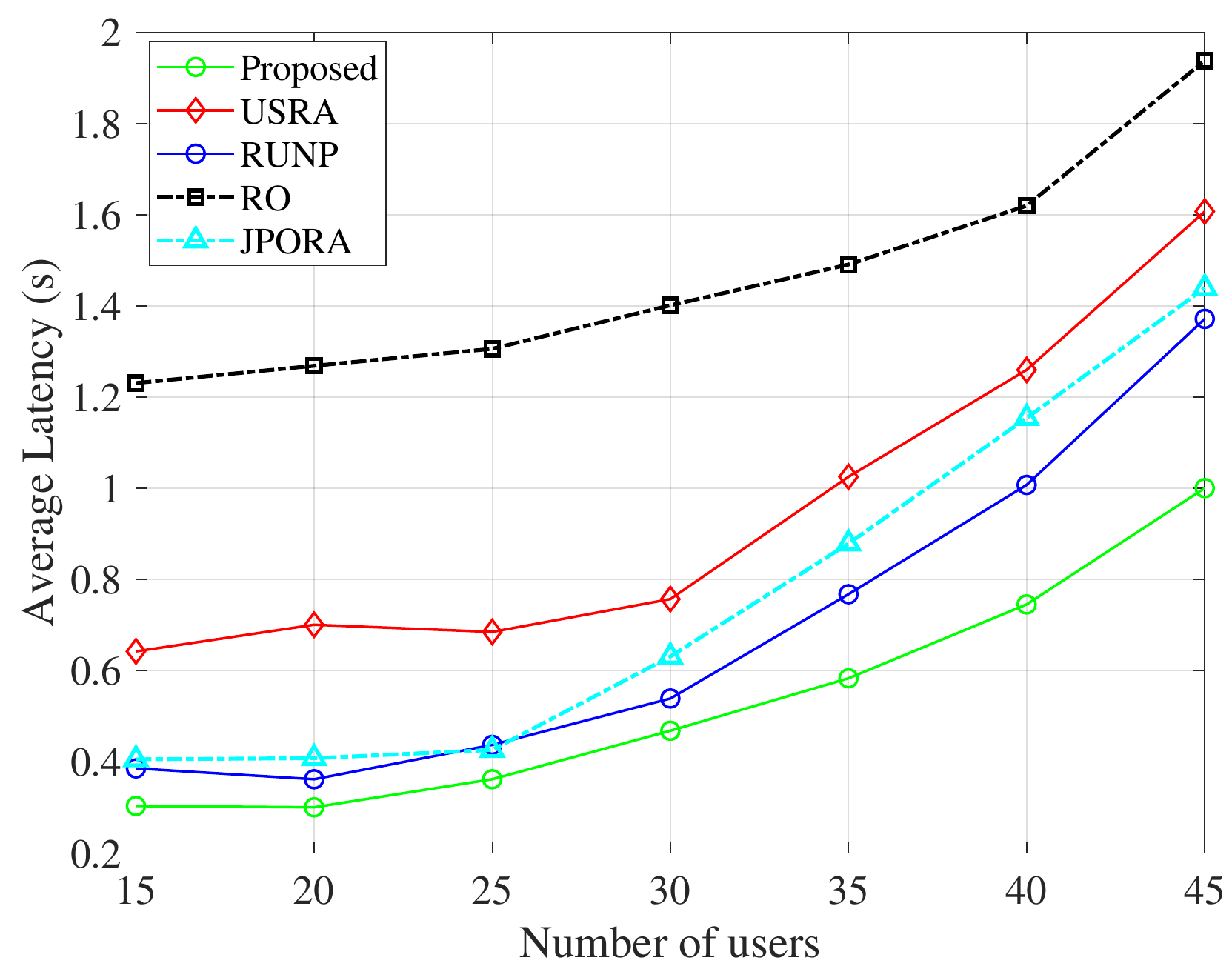}
\end{center}
\caption{Average latency of
the
four schemes with different
numbers
of users.}
\label{fig2}
\end{figure}

In Fig.~\ref{fig2}, the number of sub-channels is
set to
30 and the MR and BS computation capacity
$f^R$ and $f^B$
are set
to
$12GHz$ and $36GHz$, respectively.
The results of varying the number of users from 15 to 45 are obtained. With more users,
resources will be more and more scarce, and a larger proportion of users can only complete the computing tasks through local
execution, which causes the five curves
to rise gradually. In addition, due to the disparity in computing capacity between local and edge servers, the latency of served users
with
RO is largely determined by
the worse local latency,
and thus, the performance is poor even
when
the resources are sufficient. Compared with RUNP and JPORA, with the full utilization of edge computing resources, power control and matching game also improves
the system
performance to a certain extent.
JPORA is greatly affected by the location of users, so it cannot flexibly determine association relationships.
Similar to Table \ref{CIS}, USRA has relatively large confidence interval at the same confidence level.

\begin{figure}[!t]
\begin{center}
\includegraphics*[width=1\columnwidth,height=2.5in]{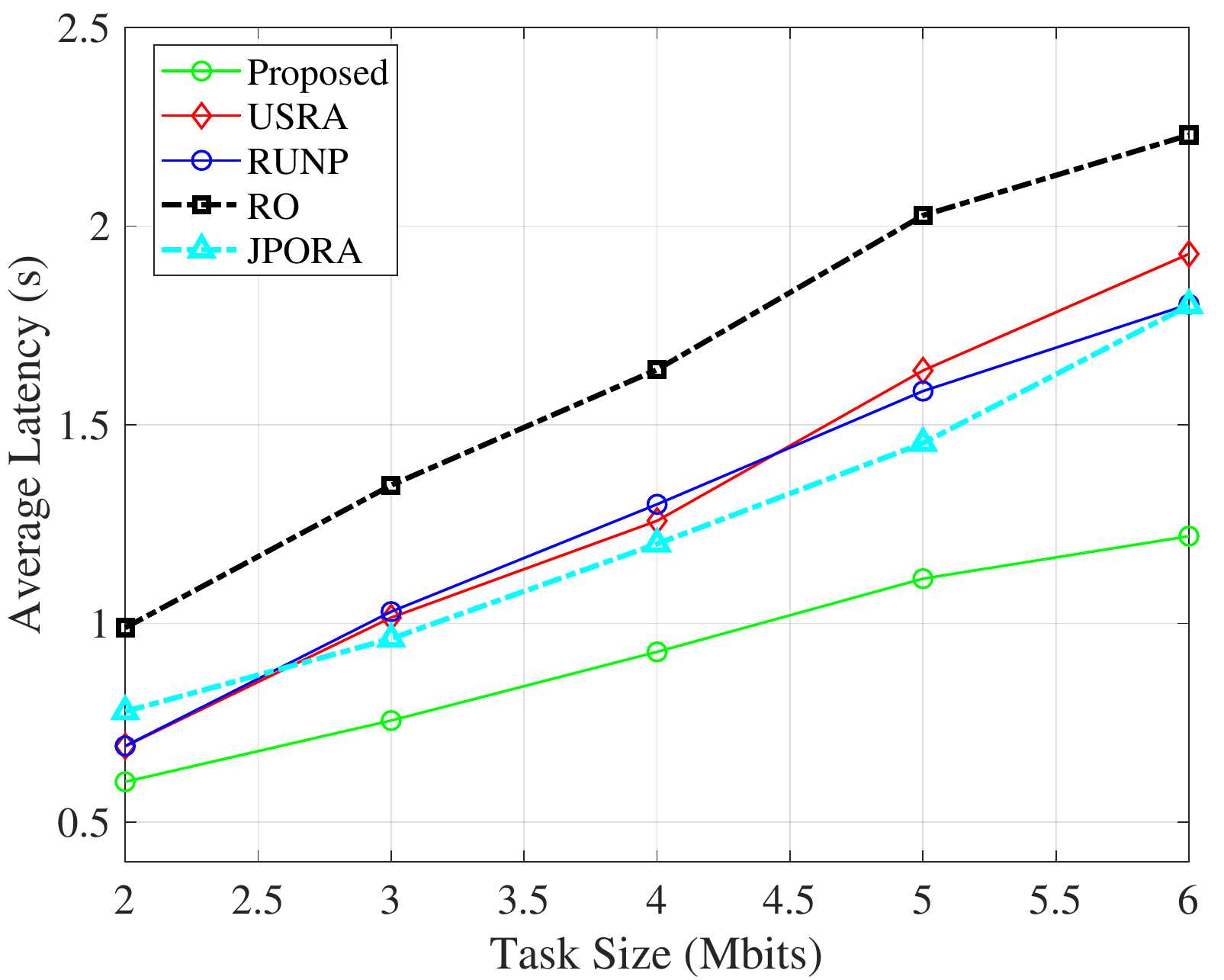}
\end{center}
\caption{Average latency of
the
four schemes with different task sizes.}
\label{fig3}
\end{figure}

\begin{figure}[!t]
\begin{center}
\includegraphics*[width=1\columnwidth,height=2.5in]{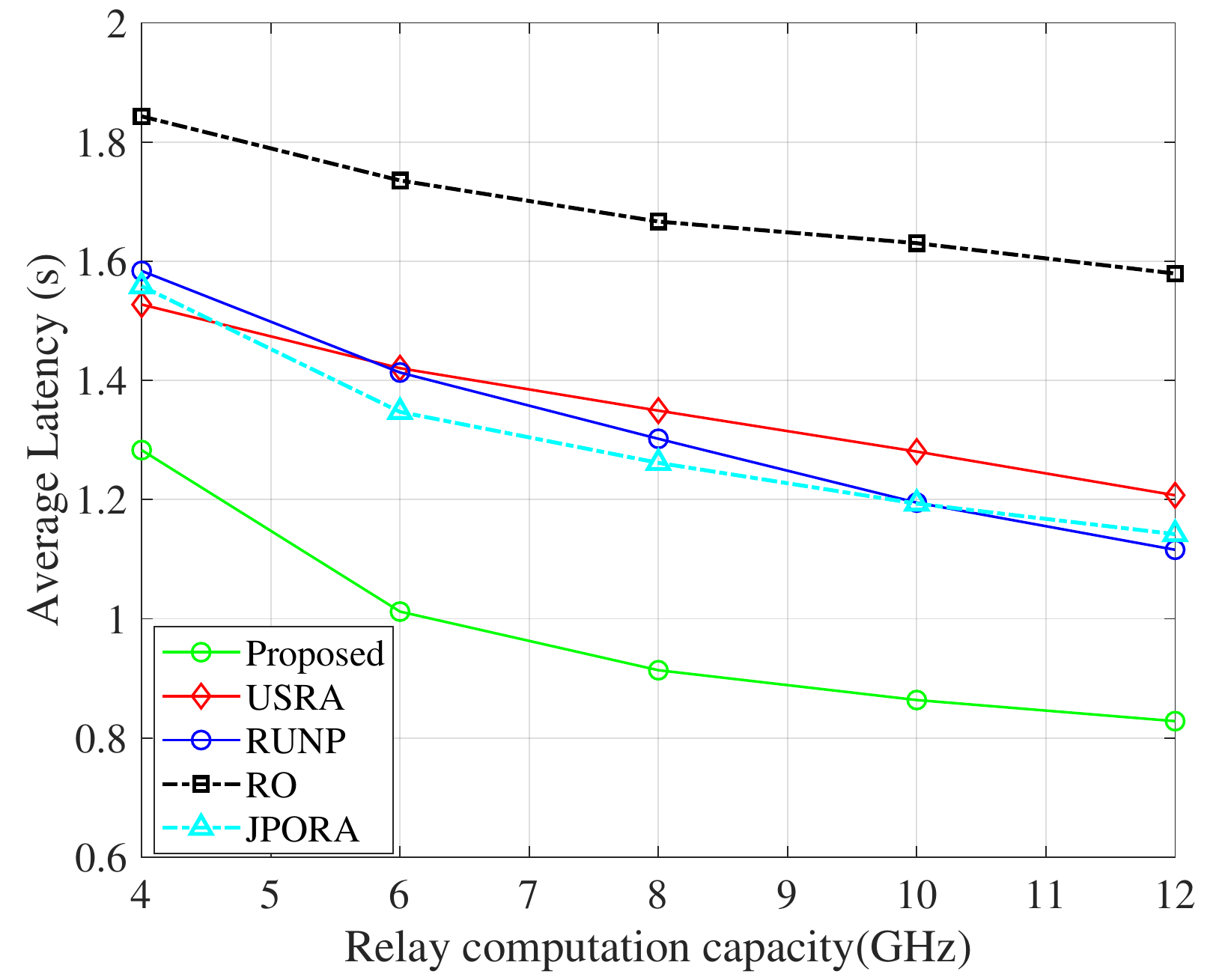}
\end{center}
\caption{Average latency of
the
four schemes with different MR computation capacities.}
\label{fig4}
\end{figure}

\begin{figure*}[!t]
\centering
\subfigure[Resource deficit: 30 Users and 20 Sub-channels ]{
\includegraphics*[width=0.95\columnwidth,height=2.5in]{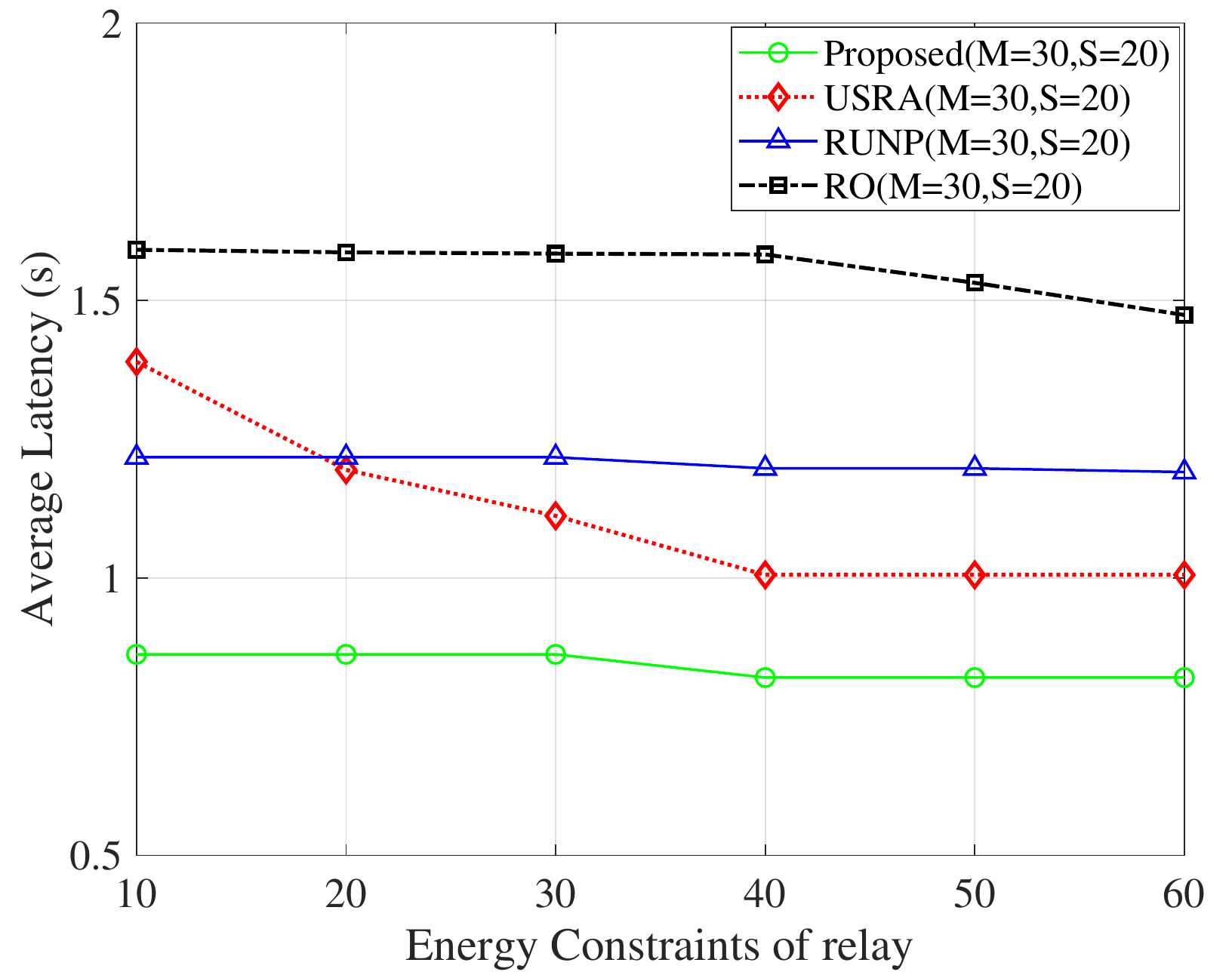}
}
\quad
\subfigure[Resource deficit: 30 Users, and 20 Sub-channels]{
\includegraphics*[width=0.95\columnwidth,height=2.5in]{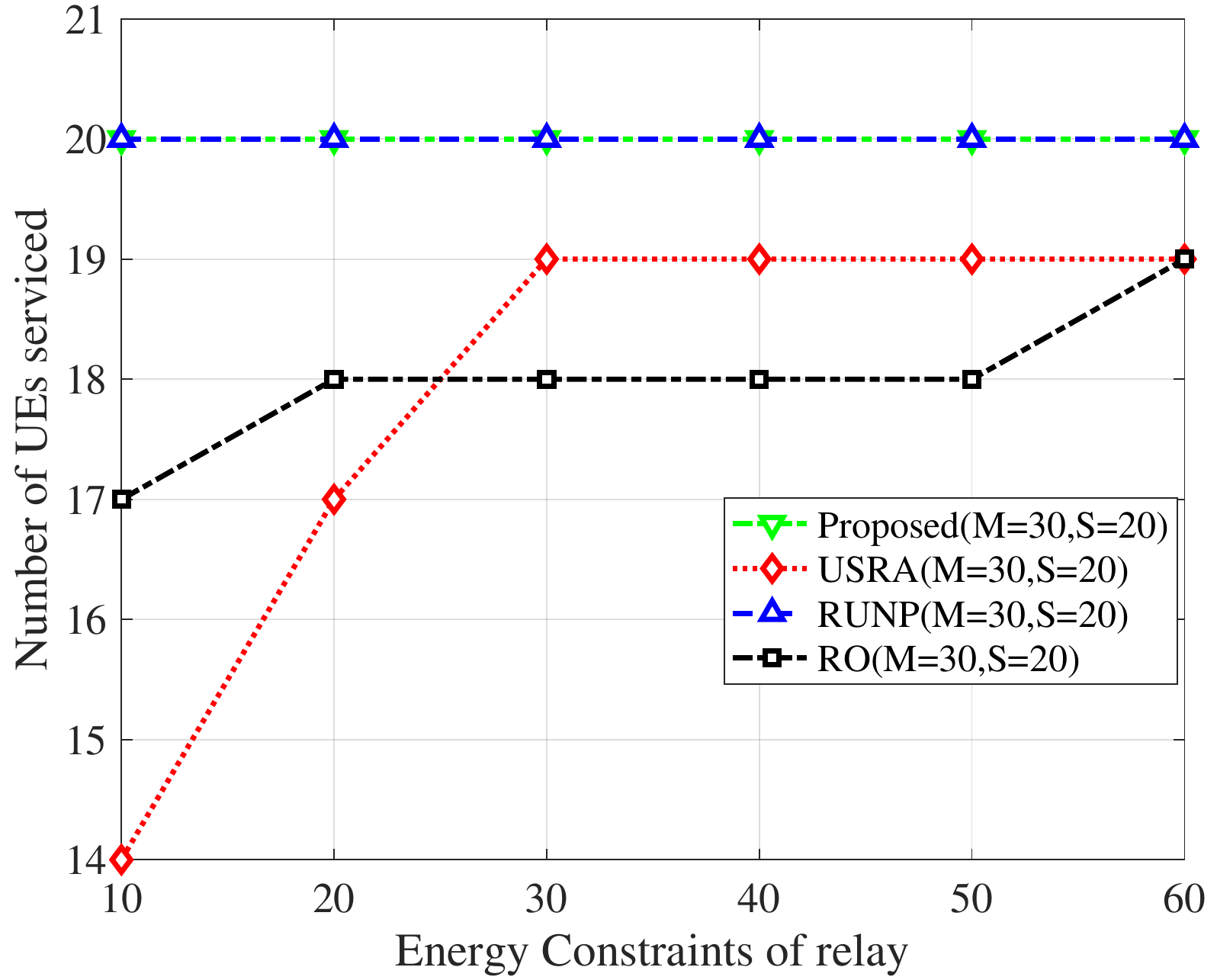}
}
\quad
\subfigure[Resource surplus with 20 Users and 25 Sub-channels]{
\includegraphics*[width=0.95\columnwidth,height=2.5in]{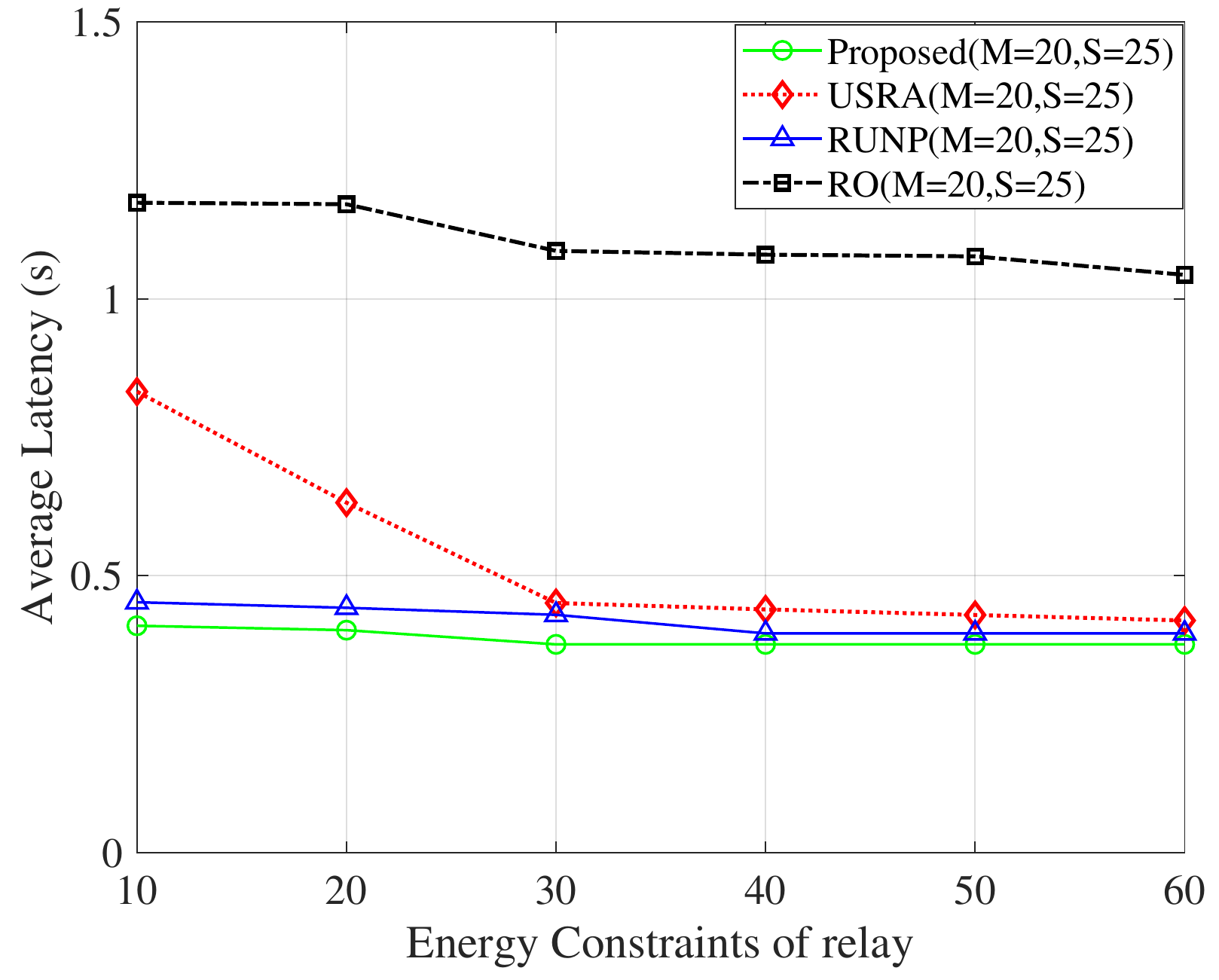}\label{3}
}
\quad
\subfigure[Resource surplus with 20 Users and 25 Sub-channels]{
\includegraphics*[width=0.95\columnwidth,height=2.5in]{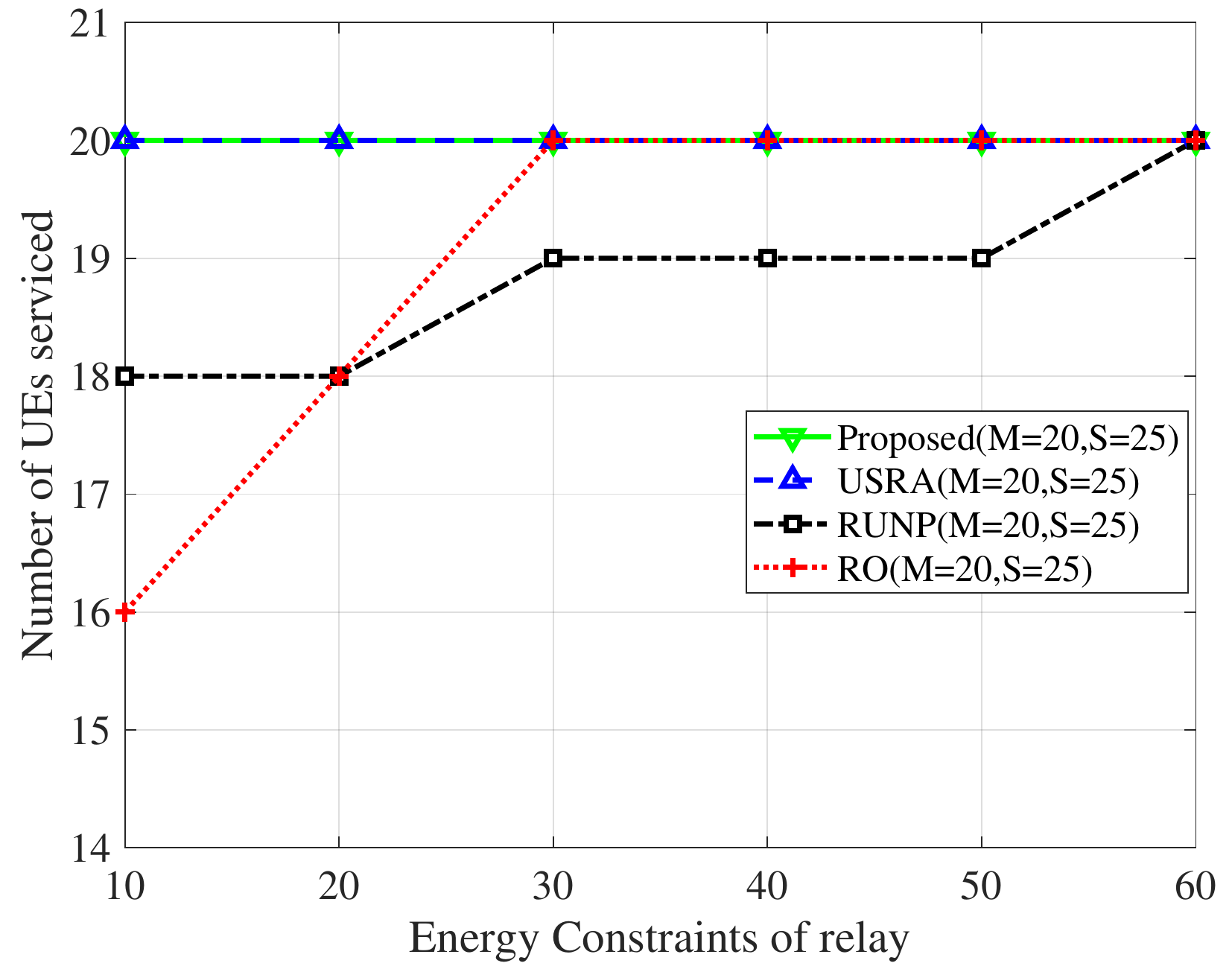}\label{4}
}
\caption{Average latency and the number of served users of
the
four schemes with different MR energy constraints.}
\label{figE}
\end{figure*}
To examine the impact of different computation task sizes of diverse user applications,
we verify the performance of the proposed algorithm
by
increasing
the
task size in Fig.~\ref{fig3}.
The number of sub-channels is
set to 20 and the number of users is 30. Therefore, users compete for limited communication resources.
For each data point,
the task size is uniformly distributed between $1$ Mbits and the task size value on that data point. Fig.~\ref{fig3} plots the  average latency, and shows that it gradually increases with
increased
data size for all the schemes. This can be explained as, with the increase in task size,
a
higher fraction of data
will be
offloaded for remote execution
to overcome
the
limited
computation capacity and local energy constraint at user devices,
which in turn leads to higher transmission and computation delay.
Furthermore,
the remote execution time increases when the offloaded task size is increased.
Our comparison study also
shows that,
the proposed scheme
has
the
lowest increasing rate. Moreover, USRA, RUNP and JPORA have similar performance,
it reduces the average latency by approximately 33\%
over
USRA.
These results show
that our algorithm has advantages for computation intensive applications.

In order to analyze the impact of edge computation resources on the
delay
performance, we plot the average latency against computation capacity of the MR in Fig.~\ref{fig4}, while maintaining
the computing capacity of
the BS
as 3 times
of the MR's computing capacity.
The number of sub-channels is set to 20 and the number of users is 30.
With the increase of computing resources, the average delay of
all schemes 
exhibits
a decreasing trend.
The performance of the proposed algorithm is not significantly improved when the computing resources
exceeds
a certain threshold, mainly because the delay of unserved users is too large
that computing resources are no longer the primary constraint.
For USRA, RUNP and JPORA, sometimes the benefit of computation resource allocation is better than that of
optimizing the served
set of
users,
and sometimes it is the opposite, which is why the intersection occurs.
The proposed
scheme achieves
35\%, 32\%, 50\% and 31\% lower latency than the other three baseline schemes in statistical average sense,  respectively.

In Fig.~\ref{figE}, we provide a comparison of
the
average delay and the number of served users performance under different MR energy constraints. In addition, we
examine
two cases: (i) resource deficit in Fig.~\ref{figE}(a)(b)
with 30 Users
and 20 Sub-channels,  and
(ii) resource surplus in Fig.~\ref{figE}(c)(d)
with 20 Users
and 25 Sub-channels.
It is obvious that
no matter it is
resource deficit or resource surplus, RUNP and the proposed
scheme
that
use
Algorithm~\ref{alg2}
are basically stable in terms of average delay,
while the delay of USRA and RO using
the
greedy algorithm (i.e., the first part of Algorithm~\ref{alg2})
decreases with the relaxation of
the constraint.
However, the performance does not improve any further beyond a certain threshold as the MR energy does not remain the dominant constraint.
In addition, RUNP and the proposed scheme
also serve more users than USRA and RO under the same conditions.
The proposed algorithm is better than
the
other algorithms in both the latency and the number of served users, especially when the resources are insufficient where there are users competing for resources.

\section{CONCLUSION AND FUTURE WORK}\label{S8}

In this paper, we
investigated the problem of user association, resource allocation (including communication and computation resources), and computation offloading in
the
uplink train-ground communication scenario.
In terms of minimizing
the
parallel computing latency under MR energy consumption control, we decomposed
the optimization problem into three sub-problems for
ease of
analysis.
The continuous variables, including the local device computational speed, offloading ratio, and transmit power of the MR were obtained through functional analysis.
Efficient resource allocation
was implemented as binary variable determined by a
dynamic matching game.
The
RACO algorithm
was proposed
to solve the
sub-problems alternately,
to obtain suboptimal solutions.
Then a heuristic MR energy consumption control algorithm
was
proposed to finally adjust the offloading rate and MR calculation frequency assigned to the served users,
in order to enforce the MR energy consumption constraint.
Through extensive simulations
under
different network parameters, we
demonstrated the superiority of the proposed scheme over three baseline schemes.
The proposed algorithm is suitable for computation offloading scenarios where the energy consumption and computing resources of edge servers are limited, but only consider the task can be completed when the users shift from the connecting BS to another adjacent BS.
Future work is to investigate the blockage problem and cooperative D2D communications
in the proposed framework.

\bibliographystyle{IEEEtran}

\vfill

\end{document}